\documentclass{aa}

\usepackage{newtxtext,newtxmath}
\usepackage[T1]{fontenc}
\usepackage{graphicx}
\usepackage{amsmath}
\usepackage{xcolor}
\definecolor{dkblue}{RGB}{54, 86, 169}

\usepackage[]{hyperref}
\usepackage{longtable}
\usepackage{pdflscape}
\usepackage{caption}
\usepackage{comment}

\newcommand{\TAB}[1] {Table~\ref{#1} }
\newcommand{\C}{\textsf{\textbf{C}}}
\newcommand{\M}{\textsf{\textbf{M}}}

\newcommand{\CW}{\textsf{\textbf{C}}_{\rm WN}}
\newcommand{\EFAC}{{\rm EFAC}}
\newcommand{\EQD}{{\rm EQUAD}}
\newcommand{\ECR}{{\rm ECORR}}

\usepackage{xcolor}
\usepackage[normalem]{ulem}

\begin{document}

\title{The Chinese Pulsar Timing Array data release I}
\subtitle{Single pulsar noise analysis}
\titlerunning{CPTA DR1: Pulsar noise}

\author{
Siyuan Chen\inst{\ref{shao},\ref{radio}}\thanks{E-mail: siyuan.chen@shao.ac.cn},
Heng Xu\inst{\ref{naoc}}\thanks{E-mail: hengxu@bao.ac.cn},
Yanjun Guo\inst{\ref{naoc},\ref{radio}}\thanks{E-mail: guoyj@bao.ac.cn},
Bojun Wang\inst{\ref{naoc}},
R. Nicolas Caballero\inst{\ref{kiaa},\ref{hou}},
Jinchen Jiang\inst{\ref{naoc}},
Jiangwei Xu\inst{\ref{pku},\ref{kiaa},\ref{naoc}},
Zihan Xue\inst{\ref{pku},\ref{kiaa},\ref{naoc}},
Kejia Lee\inst{\ref{pku},\ref{naoc},\ref{radio},\ref{laser}},
Jianping Yuan\inst{\ref{xao}},
Yonghua Xu\inst{\ref{ynao}},
Jingbo Wang\inst{\ref{lishui}},
Longfei Hao\inst{\ref{ynao}},
Jintao Luo\inst{\ref{time}},
Jinlin Han\inst{\ref{naoc}},
Peng Jiang\inst{\ref{naoc}},
Zhiqiang Shen\inst{\ref{shao}}
Min Wang\inst{\ref{ynao}},
Na Wang\inst{\ref{xao}},
Renxin Xu\inst{\ref{kiaa},\ref{pku},\ref{nuclear}},
Xiangping Wu\inst{\ref{naoc}},
Lei Qian\inst{\ref{naoc}},
Xin Guan\inst{\ref{naoc}},
Menglin Huang\inst{\ref{naoc}},
Chun Sun\inst{\ref{naoc}}
and Yan Zhu\inst{\ref{naoc}}.
}
\authorrunning{Chen S. et al.}

\institute{
{Shanghai Astronomical Observatory, Chinese Academy of Sciences, Shanghai 200030, P.~R.~China\label{shao}}\and
{Key Laboratory of Radio Astronomy and Technology, Chinese Academy of Sciences, Beijing 100101, P.~R.~China\label{radio}}\and
{National Astronomical Observatories, Chinese Academy of Sciences, Beijing 100101, P.~R.~China\label{naoc}}\and
{Kavli Institute for Astronomy and Astrophysics, Peking University, Beijing 100871, P.~R.~China\label{kiaa}}\and
{Hellenic Open University, School of Science and Technology, 26335 Patras, Greece\label{hou}}\and
{Department of Astronomy, School of Physics, Peking University, Beijing 100871, P.~R.~China\label{pku}}\and
{Beijing Laser Acceleration Innovation Center, Huairou, Beijing 101400, P.~R.~China\label{laser}}\and
{Xinjiang Astronomical Observatory, Chinese Academy of Sciences, Urumqi 830011, Xinjiang, P.~R.~China\label{xao}}\and
{Yunnan Astronomical Observatories, Chinese Academy of Sciences, Kunming 650216, Yunnan, P.~R.~China\label{ynao}}\and
{Institute of Optoelectronic Technology, Lishui University, Lishui, Zhejiang, 323000, P.~R.~China\label{lishui}}\and
{National Time Service Center, Chinese Academy Of Sciences, Xi'an 710600, P.~R.~China\label{time}}\and
{State Key Laboratory of Nuclear Physics and Technology, School of Physics, Peking University, Beijing 100871, P.~R.~China\label{nuclear}}
}

\date{Accepted XXX. Received YYY; in original form ZZZ}

\abstract 
{
The Chinese Pulsar Timing Array (CPTA) has collected observations from 57 millisecond pulsars using the Five-hundred-meter Aperture Spherical Radio Telescope (FAST) for close to three years, for the purpose of searching for gravitational waves (GWs). To robustly search for ultra-low-frequency GWs, pulsar timing arrays (PTAs) need to use models to describe the noise from the individual pulsars. We report on the results from the single pulsar noise analysis of the CPTA data release I (DR1). Conventionally, power laws in the frequency domain are used to describe pulsar red noise and dispersion measurement (DM) variations over time. Employing Bayesian methods, we found the choice of number and range of frequency bins with the highest evidence for each pulsar individually. A comparison between a dataset using DM piecewise measured (DMX) values and a power-law Gaussian process to describe the DM variations shows strong Bayesian evidence in favour of the power-law model. Furthermore, we demonstrate that the constraints obtained from four independent software packages are very consistent with each other. The short time span of the CPTA DR1, paired with the large sensitivity of FAST, has proved to be a challenge for the conventional noise model using a power law. This mainly shows in the difficulty to separate different noise terms due to their covariances with each other. Nineteen pulsars are found to display covariances between the short-term white noise and long-term red and DM noise. With future CPTA datasets, we expect that the degeneracy can be broken. Finally, we compared the CPTA DR1 results against the noise properties found by other PTA collaborations. While we can see broad agreement, there is some tension between different PTA datasets for some of the overlapping pulsars. This could be due to the differences in the methods used to obtain the constraints or the different frequency range that is probed with the CPTA DR1, which probes a higher frequency range compared to the other PTAs.
}


\maketitle

\section{Introduction}
\label{sect:intro}

Pulsar timing array \citep[PTA,][]{fb1990} experiments aim to detect ultra-low-frequency gravitational waves (GWs) in the nano- to milli-Hertz regime, where supermassive black hole binaries (SMBHBs) are a prime target, but cosmological sources are also possible. The Chinese Pulsar Timing Array \citep[CPTA,][]{lee2016} is a collaboration of various Chinese institutions using the Five-hundred-meter Aperture Spherical Telescope \citep[FAST,][]{jyg+2019} as the main instrument to observe millisecond pulsars with the highest single-dish sensitivity to date\footnote{The FAST data policy and instructions on how to request public data can be found online: \url{https://fast.bao.ac.cn}.}.

The detection of any GW signal with PTA data is challenging as it needs to be found amongst numerous other possible sources of noise. For a gravitational wave background (GWB), e.g. from a population of SMBHBs or from cosmological sources, the characteristic spatial correlation of the signal across a network of pulsars \citep[][henceforth referred to as HD]{hd1983} acts as a smoking gun. For individual GW sources, like a single binary, the resulting GW signal is deterministic. The signals from both, the GWB and individual sources, can be compared against a noise-only model.
To be able to distinguish GW signals from pulsar related noise terms, robust noise models for each pulsar need to be employed in the analysis. \cite{grs+2021} and \cite{cbp+2022} have shown improvements in the detection significance of a GWB signal in PTA datasets using optimized noise models.

Globally, the efforts are coordinated in the International Pulsar Timing Array \citep[IPTA,][]{ipta_dr1,ipta_dr2} consortium, with regional member PTAs operating in Europe with the European Pulsar Timing Array \citep[EPTA,][]{epta2013}, North America with the North American Nanohertz Observatory for Gravitational waves \citep[NANOGrav,][]{nanograv2013}, Australia with the Parkes Pulsar Timing Array \citep[PPTA,][]{ppta2013}, and India with the Indian Pulsar Timing Array \citep[InPTA,][]{inpta_dr1}. Collaborations also exist in South Africa with the MeerKAT PTA \citep{mpta_dr1} and in China with the CPTA. A series of coordinated papers searching for a GWB have been published by the member PTAs of the IPTA \citep{ng15_gwb,eptadr2_gwb,pptadr3_gwb} and the CPTA \citep{cpta_dr1} based on their most recent datasets. The MeerKAT PTA has also found evidence of a GW signal in its dataset \citep{meerkat_gwb}.

Different PTA collaborations have applied different methods to set and optimize the noise models used for their analysis \citep{eptadr2_noise,pptadr3_noise,ng15_noise,meerkat_data}. Using the recent datasets, they find a consistent common signal with varying support for the HD correlation, which indicates some evidence of the GW origin of the measured signal. A detailed comparison between the NANOGrav, EPTA+InPTA, and PPTA results was performed by the IPTA \citep{ipta_comp}.

This paper is part of a series using the CPTA data release I (DR1) and expands upon different aspects of the overview work in \cite{cpta_dr1}. Here, we analyse the single pulsar noise properties of 57 pulsars that form the CPTA DR1 with time spans up to 3.4 years and a weekly cadence (for details on the dataset see Xu et al., in prep). Section 2 provides a summary of the analysis methods and different software packages used. The results of the analysis are presented in Section 3. We discuss comparisons with noise properties from other PTAs and conclude in Section 4.

\section{Methods}
\subsection{Analysis}
\label{sec:methods}

A more detailed description of the overall likelihood for the single pulsar noise model can be found in \cite{ccg+2021} and \cite{cbp+2022}, and references therein. The likelihood function with units of $s^{-n}$ can be written as
\begin{equation}
\label{eq:psrlik}
    L_{\rm{PSR}} = \frac{e^{-\frac{1}{2}(\delta t_{\rm post})^{T}\C^{-1}(\delta t_{\rm post})}}{\sqrt{(2\pi)^n|\C|}} \,,
\end{equation}
where the vector of the timing post-fit residuals, $\delta t_{\rm post}$ , and vector of the timing pre-fit residuals, $\delta t_{\rm pre}$ , have $n$ elements with corresponding observation time, $t_j$, uncertainty, $\sigma_j$ , and radio frequency, $\nu_j$. They are related via the design matrix, $\M,$ of the timing model parameters, $\xi$ \citep{tempo2}: $\delta t_{\rm post} = \delta t_{\rm pre} - \M\xi$. The timing model parameters can be analytically marginalized \citep{vhl+2009,vhl2013,vhv2014} \footnote{Note that we marginalize over $\M$ by employing the Gaussian process description of the timing model and using an `infinite` prior on $\xi$ \citep{ng9,tlb+2017}.}. The noise covariance matrix, $\C,$ can be split into three components, white noise (WN), red noise (RN), and dispersion measure (DM) variation of the pulsar: $\C = \CW + \C_{\rm RN} + \C_{\rm DM}$.

Three parameters are applied to the initial estimate on the uncertainty, $\sigma_j$, of the times of arrival (TOAs) from the template matching method to test how well we can understand the white noise of the observing system and pulsar intrinsic properties. They are called EFAC, a multiplicative factor, EQUAD, an additional quadratic term, and ECORR, a white noise correlated between TOAs from a single observation. Together they result in the white noise covariance matrix
\begin{equation}
    {\rm C}_{{\rm WN}, jk} = \EFAC^2 \sigma_j \sigma_k \delta_{jk} + \EQD^2 \delta_{jk} + \ECR^2 {\rm C}_{\ECR,jk} \,,
\end{equation}
where $\delta_{jk}$ is the Kronecker delta symbol and the ECORR covariance matrix element ${\rm C}_{\ECR,jk} = 1$, when the $j$-th and the $k$-th data points are of the same observing time but of different radio frequencies; and matrix element ${\rm C}_{\ECR,jk} = 0$ otherwise. We note that ECORR is required for the CPTA data, as we produce channelized TOAs. Thus, pulsar jitter can introduce white noise that is correlated between the TOAs of the same epoch (i.e. between different radio frequencies).
Pulsar jitter noise should decrease as the observation time is increased averaging over more rotations. We tested this hypothesis by comparing the models of EQUAD and ECORR, whose values depend on the observation time length against the model where they do not. We find that most pulsars do not favour either model, thus we used the conventional time-independent model for the main analysis. More details can be found in Appendix~\ref{sec:jitter}.

The average $\langle \, \rangle$ of many realizations of the timing residuals can be described by a stochastic process. Using the Wiener-Khinchin theorem, the covariance, ${\rm C}_{{\rm X}, jk}$ , between two elements of the residual time series $\delta t_{{\rm X}, j}$ and $\delta t_{{\rm X}, k}$ at times $t_j$ and $t_k$ with radio frequencies $\nu_j$ and $\nu_k$ can be defined using a one-sided ($f>0$) power spectral density (PSD) $S_{\rm X}(f,\nu)$:
\begin{equation}
{\rm C}_{{\rm X}, jk} = \langle \delta t_{{\rm X}, j} \, \delta t_{{\rm X}, k} \rangle = \int_0^\infty S_{\rm X}(f,\nu) e^{2\pi i f(t_j-t_k)} df \,,
\label{eq:cov}
\end{equation}
where X can be either RN or DM and $\nu^2=\nu_j\nu_k$. For RN, the PSD is independent of the radio frequency, while for DM it is not. If the PSD is constant across all frequencies, this equation can also describe WN.

For long-term variations in the residual time series, also known as red noise, the PSD $S_{\rm RN}(f)$ is usually modelled as a power law in the Fourier frequency domain:
\begin{equation}
S_{\rm RN}(f) = \frac{A_{\rm RN}^2}{12\pi^2} \left(\frac{f}{f_{\rm yr}}\right)^{-\gamma_{\rm RN}} {\rm yr}^3\,,
\label{eq:rn}
\end{equation}
with dimensionless amplitude $A_{\rm RN}$ and spectral index $\gamma_{\rm RN}$ sampled at frequencies $f$ scaled by $f_{\rm yr} = 1/{\rm yr}$. The integral over an infinite frequency range in Eqn.~\eqref{eq:cov} can in practice be approximated by a sum over a limited number of frequency bins.

Two common methods are used to model the DM variations. DM variations can be measured in a piece-wise fashion between subsequent observations to produce a DM time series. These DM values can be applied to the residuals to subtract the effect of variations of the interstellar medium. This model, called DMX \citep{nanograv2015}, can be compared against a DM variation model that assumes a smooth variation over time with a Gaussian process (GP).
Similarly to \cite{lmh+2024}, we directly compared the results from a DM GP to those from a DMX-style data set.
The DM GP models the DM variation over time after fitting for a constant, linear, and quadratic term.

Similar to the red noise, we used a power law to model the timing residuals induced by the remaining small DM variations.
In the literature, there are two common definitions of the power-law model: \texttt{TEMPONEST} \citep[TN, ][]{lah+2014} and \texttt{ENTERPRISE} \citep[EP, ][]{enterprise}. In the TN convention, the PSD $S_{\rm DM}(f)$ is written as
\begin{equation}
S_{\rm DM}(f,\nu) = \frac{A_{\rm DM,TN}^2}{K^2\nu^4} \left(\frac{f}{f_{\rm yr}}\right)^{-\gamma_{\rm DM}} {\rm yr}^3\,,
\label{eq:dm_tn}
\end{equation}
where $A_{\rm DM,TN}$ has units of ${\rm pc\, cm^{-3}\, s^{-1}}$ and $K=2.41\times10^{-4}{\, \rm pc\, cm^{-3}\,MHz^{-2}\,s^{-1}}$ is the DM constant. The power law can also be written in the EP convention as
\begin{equation}
S_{\rm DM}(f,\nu) = \frac{A_{\rm DM,EP}^2}{12\pi^2} \left(\frac{f}{f_{\rm yr}}\right)^{-\gamma_{\rm DM}} \left(\frac{\nu}{\nu_{\rm ref}}\right)^{-4} {\rm yr}^3\,,
\label{eq:dm_ep}
\end{equation}
where $A_{\rm DM,EP}$ is dimensionless and $\nu_{\rm ref}=1400\,{\rm MHz}$ is a reference radio frequency. Comparing the two conventions gives a relation between the TN and EP amplitudes:
\begin{equation}
    A_{\rm DM,EP}=\sqrt{12\pi^2}A_{\rm DM,TN}\Big/(K\nu_{\rm ref}^2)\,.
\end{equation}
Note that in this work, we employ the TN convention and denote $A_{\rm DM,TN}=A_{\rm DM}$ in the following.

Additionally, we also model deterministic variations with a yearly period in the DM time series:
\begin{equation}
    \delta {\rm DM}_{{\rm yr}, j} = A_{\rm yr} \sin(2\pi f_{\rm yr} t_j + \phi_{\rm yr})\,,
\end{equation}
where $A_{\rm yr}$ in units of ${\rm pc\, cm^{-3}}$ is the amplitude of the yearly variations in the DM time series and $\phi_{\rm yr}$ is the phase of the yearly sinusoid. The exact value of $\phi_{\rm yr}$ depends on what time is chosen to correspond to $\phi_{\rm yr}=0$.

The DM variations induced timing residuals $\delta t_{\rm DM_{\rm Y}}$ in s and the DM time series $\delta {\rm DM}_{\rm Y}$ in ${\rm pc\, cm^{-3}}$ are related via \citep{lah+2014}
\begin{equation}
   \delta {\rm DM}_{{\rm Y}, j} = K\nu_j^2 \delta t_{{\rm DM}_{\rm Y}, j} \,,
\label{eq:knu2}
\end{equation}
with Y denoting either the GP model or yearly variations.

To robustly model the red and DM noise as power laws in an efficient way, the Fourier frequency components need to be chosen carefully. While in theory a large enough number of frequency bins should provide an accurate fit, recent works have shown that in practice restricting the number of frequency bins can limit the covariance between white and red noise \citep{cbp+2022,eptadr2_noise}. In addition, such a covariance could be falsely attributed to a GW signal, see e.g. \cite{ipta_dr2_cgw}. A small number of frequency bins will also reduce the computational cost.
The conventional choice in PTA analysis has been $1/T$, $2/T$, up to $n_{\rm high}/T$, linearly spaced frequency bins with width $1/T$, up to a maximum number $n_{\rm high}$. This approach has been shown to work well and efficiently, if there is a strong and dominant red signal in the lowest frequency bins. One can use logarithmically spaced frequency bins if the white noise is the dominant signal. A mixture between the two has been proposed by \cite{vhv2015}, which should result in an accurate model that is also efficient to evaluate computationally. However, that model requires knowledge of the transition frequency between red and white noise. For the CPTA analysis, we used $n_{\rm bin}$ logarithmically and evenly spaced frequency bins between $1/T$ and $n_{\rm high}/T$.
We applied the Romberg summation weights to the frequency bins to increase the accuracy of the model, details of which can be found in Appendix~\ref{sec:romberg}.

Following the same principle as \cite{cbp+2022}, we searched for the optimal frequency choice for the pulsar red and DM variation noise.
We also used the Bayesian method to determine the optimal largest frequency bin, $n_{\rm high}$. Additionally, we also optimized for the lowest total number of frequency bins, $n_{\rm bin}$. The Bayes factor, $\mathcal{B,}$ can be computed by comparing the evidence, $\mathcal{Z}_i$ , between two models: $\mathcal{B}_{21} = \mathcal{Z}_2 / \mathcal{Z}_1$. If $\mathcal{B}_{21} > 1,$ then model 2 is favoured over model 1, while model 1 is favoured over model 2, if $\mathcal{B}_{21} < 1$. For a scale of significance see e.g. \cite{kr1995}.

\subsection{Software}
\label{sec:code}

We used the \texttt{TEMPO2} \citep{tempo2} package for the timing model design matrix and the residuals. To compute the likelihood function we employed four different software packages and samplers for the Bayesian analysis.

\texttt{TEMPONEST} \citep{lah+2014} is a Bayesian pulsar timing analysis software package written in the C programming language, which employs a nested-sampling approach \citep{fh2008}. As an integrated plugin of TEMPO2, it models the red noise and DM variation using either a power-law model or a model-independent method. Within our analyses, we employed \texttt{MULTINEST} \citep{fh2008} as the sampler.

\texttt{ENTERPRISE} \citep{enterprise} is a suite of python-based scripts that can be used with the timing model fit and residuals from \texttt{TEMPO2} using \texttt{libstempo} \citep{libstempo}. Signals and noises are modelled as Gaussian processes, typically with power-law spectra \citep{vhv2014}. The likelihood computation from \texttt{ENTERPRISE} is passed to \texttt{PTMCMCSAMPLER} \citep{ptmcmc} for the Bayesian parameter estimation via Monte Carlo Markov chain sampling.

\texttt{42} \citep{cll+2016} is a software package written in python for Bayesian PTA data analysis. The red noise and DM terms can be modelled with power-law spectra or other models. The sampler \texttt{MULTINEST} \citep{fh2008} is used for the single pulsar noise analysis.

\texttt{42++} is a recently developed Bayesian PTA data analysis software written in the C++ language. It relies on the \texttt{intel OneAPI} package to implement high performance numerical linear algebra operations. The software package supports Metropolis–Hastings and parallel tempering sampling.

\begin{table}
\renewcommand{\arraystretch}{1.3}
\caption{Prior choice for the parameters in the Bayesian analysis.}
\centering
\begin{tabular}{c|c|c}
\hline\hline 
Parameter  &  Prior distribution  &  Prior range  \\
\hline
EFAC  &  Uniform  &  $[0.1,5]$  \\
EQUAD  &  Log10-Uniform  &  $[-9,-5]$  \\
ECORR  &  Log10-Uniform  &  $[-9,-5]$  \\
$A_{\rm RN}$, $A_{\rm DM}$  &  Log10-Uniform  &  $[-18,-10]$  \\
$A_{\rm c,RN}$  &  Log10-Uniform  &  $[-9,-5]$  \\
$A_{\rm c,DM}$  &  Log10-Uniform  &  $[-6,-2]$  \\
$\gamma_{\rm RN}$, $\gamma_{\rm DM}$  &  Uniform  &  $[0,7]$  \\
$\alpha_{\rm RN}$, $\alpha_{\rm DM}$  &  Uniform  &  $[-0.5,4]$  \\
$n_{\rm bin}$  &  Discrete values  &  $[33,65,129,257,513]$  \\
$n_{\rm high}$  &  Log10-Uniform  &  $[0,3]$  \\
$A_{\rm yr}$  &  Log10-Uniform  &  $[-6,-2]$  \\
$\phi_{\rm yr}$  &  Uniform  &  $[0,2\pi]$  \\
\hline\hline
\end{tabular}
\label{tab:prior}
\end{table}

The prior choice for the parameters can be found in \TAB{tab:prior}. For 42++, different priors for the power-law processes are adopted. Instead of the spectral amplitude $A_{\rm X}$ and index $\gamma_{\rm X}$, 42++ uses the characteristic amplitude $A_{\rm c,X}$ and index $\alpha_{\rm X}$ to specify the power-law model, such that the PSD becomes
\begin{equation}
S_{\rm X}(f)= \frac{A_{\rm c,X}^2}{f} \left(\frac{f}{f_{\rm yr}}\right)^{-2\alpha_{\rm X}}\,.
\label{eq:42pp}
\end{equation}
The characteristic amplitude $A_{\rm c,X}$ has the intuitive physical meaning of the strength of the signal at a yearly timescale. For red noise, $A_{\rm c,RN}$ has the same unit as the timing residuals, i.e. second, while for DM, $A_{\rm c,DM}$ has the DM unit, i.e. $\rm pc\, cm^{-3}$.

The relations between the different amplitudes can be derived from comparing the PSDs from Eq.~\eqref{eq:rn} with Eq.~\eqref{eq:42pp}, and Eq.~\eqref{eq:dm_tn} with Eq.~\eqref{eq:42pp} by setting $\gamma_{\rm X} = 2\alpha_{\rm X} + 1$ for both RN and DM \footnote{The \texttt{42++} PSD for the DM describes the DM time series $\delta {\rm DM}$, thus Eq.~\eqref{eq:knu2} needs to be applied to Eq.~\eqref{eq:dm_tn} before the comparison.}:
\begin{equation}
    A_{\rm c,RN} = A_{\rm RN} \, {\rm yr} \Big/ \sqrt{12\pi^2} \,,
    A_{\rm c,DM} = A_{\rm DM} \, {\rm yr} \,.
\end{equation}

\section{Results}

The noise optimization and selection was done in three steps:
\begin{enumerate}
\item The choice of the number of frequency bins and the high frequency cutoff was determined in a Bayesian model comparison.
\item The single pulsar noise analysis was run with these choices with our four software packages.
\item The final noise terms were selected based on the highest Bayes factor.
\end{enumerate}
The optimal frequency bin choice is given in \TAB{tab:freq}, where $n_{\rm bin}$ is the number of logarithmically and evenly spaced bins between $f_0 = 1/T$ and $f_{\rm high} = n_{\rm high}/T$ , with $T$ being the time span from the earliest to the latest TOA of a given pulsar's dataset. In our optimization and selection processes, we modelled the high-frequency cutoff as one of the noise model parameters. We then ran the Bayesian inference in step 1 using \texttt{42++} only with different choices of number of frequency bins, where we tried 33, 65, 129, 257, and 513 (i.e. $2^n+1$, where $n$ is the order of the integration). These numbers satisfy the requirement of the Romberg integration. For all runs, we note that the values of the Bayesian evidence are nearly the same if the number of frequency bins is large enough. To reduce the computational cost, we chose the minimal number of frequency bins, at which the Bayesian evidence is `not significantly worse' than the best run with the highest Bayesian evidence. Here, `not significantly worse' means that the difference between the logarithmic Bayesian evidence is less than 1.
The choice was applied to both the red and DM noise. In most cases, a small number is sufficient to describe pulsars without red or DM noise. Pulsars with noticeable red or DM noise need in general a slightly large number for $n_{\rm high}$ to reach into the high frequencies. In most pulsars, 33 bins are enough to describe the power law sufficiently well. A few pulsars favour the inclusion of up to 129 frequency bins.

\begin{figure*}
\centering
\includegraphics[width=0.95\linewidth]{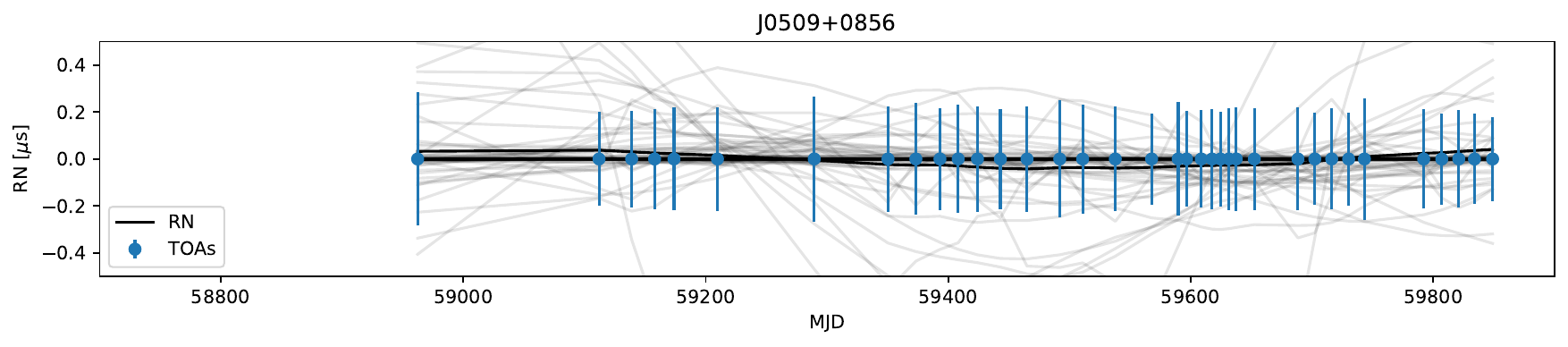}
\includegraphics[width=0.95\linewidth]{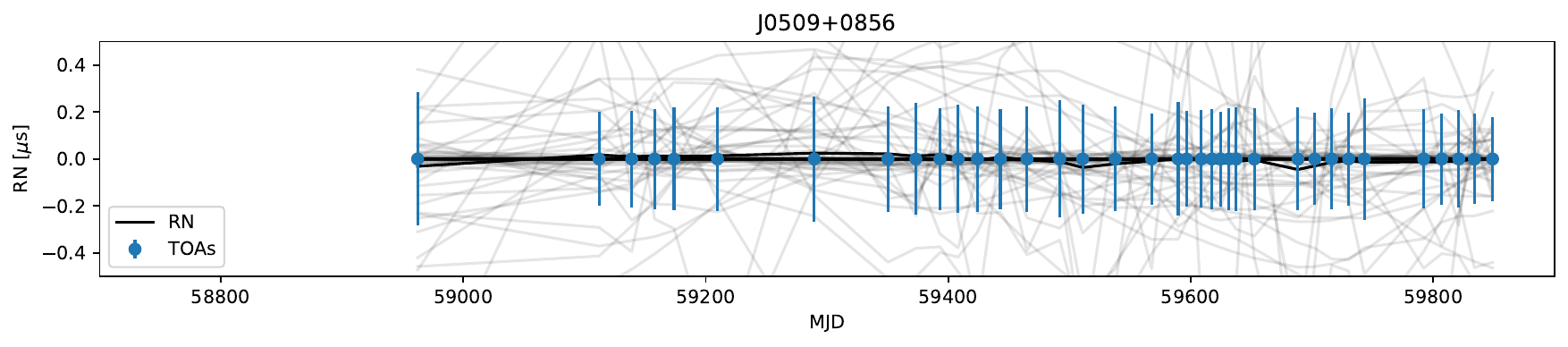}
\includegraphics[width=0.95\linewidth]{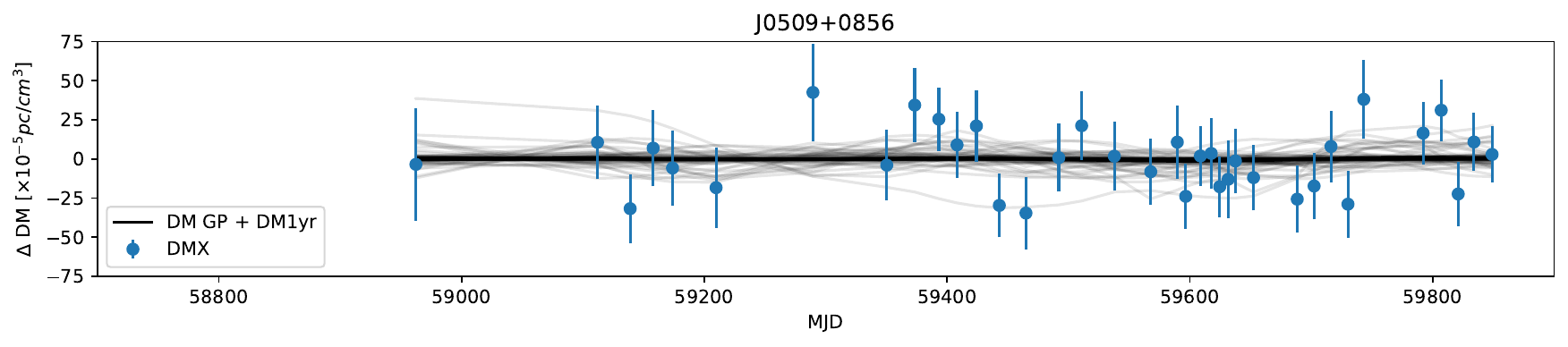}

\includegraphics[width=0.95\linewidth]{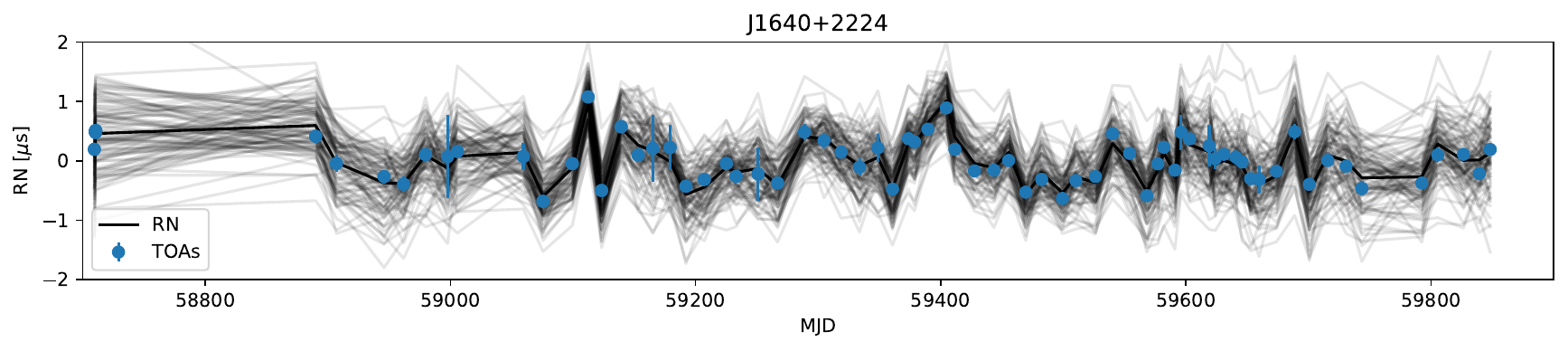}
\includegraphics[width=0.95\linewidth]{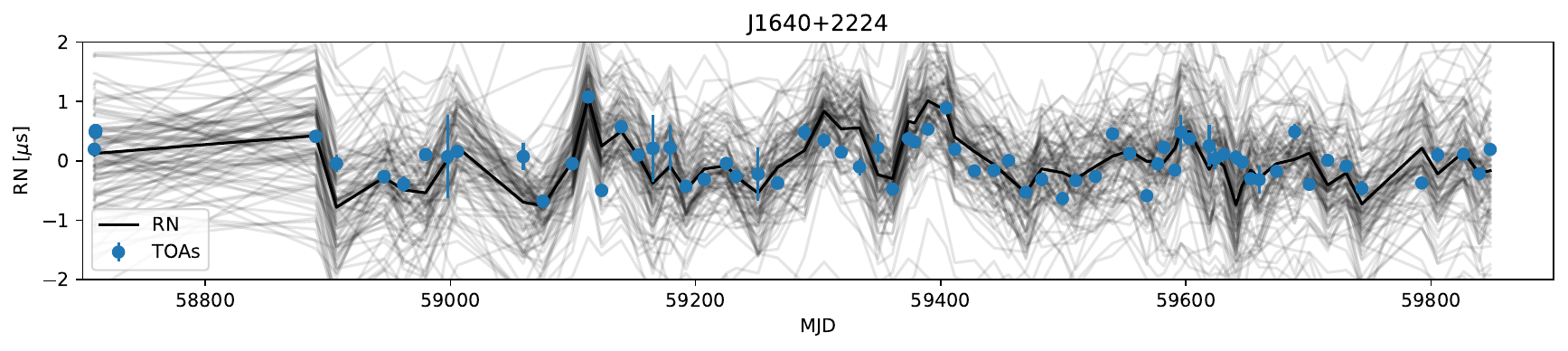}
\includegraphics[width=0.95\linewidth]{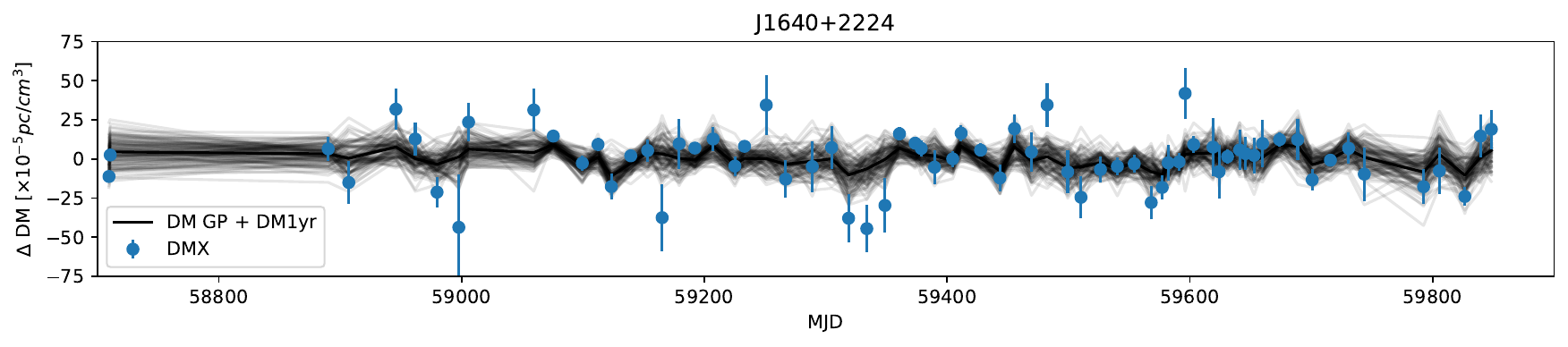}

\caption{Time series reconstructions for J0509+0856 in the top three and J1640+2224 in the bottom three panels. From
top to bottom, each set of three panels shows  the red noise for the DM GP and DMX datasets, as well as the DM GP reconstructions compared the DMX values (where linear and quadratic trends have been removed). The blue points indicate the median residual values and $1\sigma$ uncertainties.}
\label{fig:gp_series}
\end{figure*}

\begin{figure*}
\centering
\includegraphics[width=0.9\linewidth]{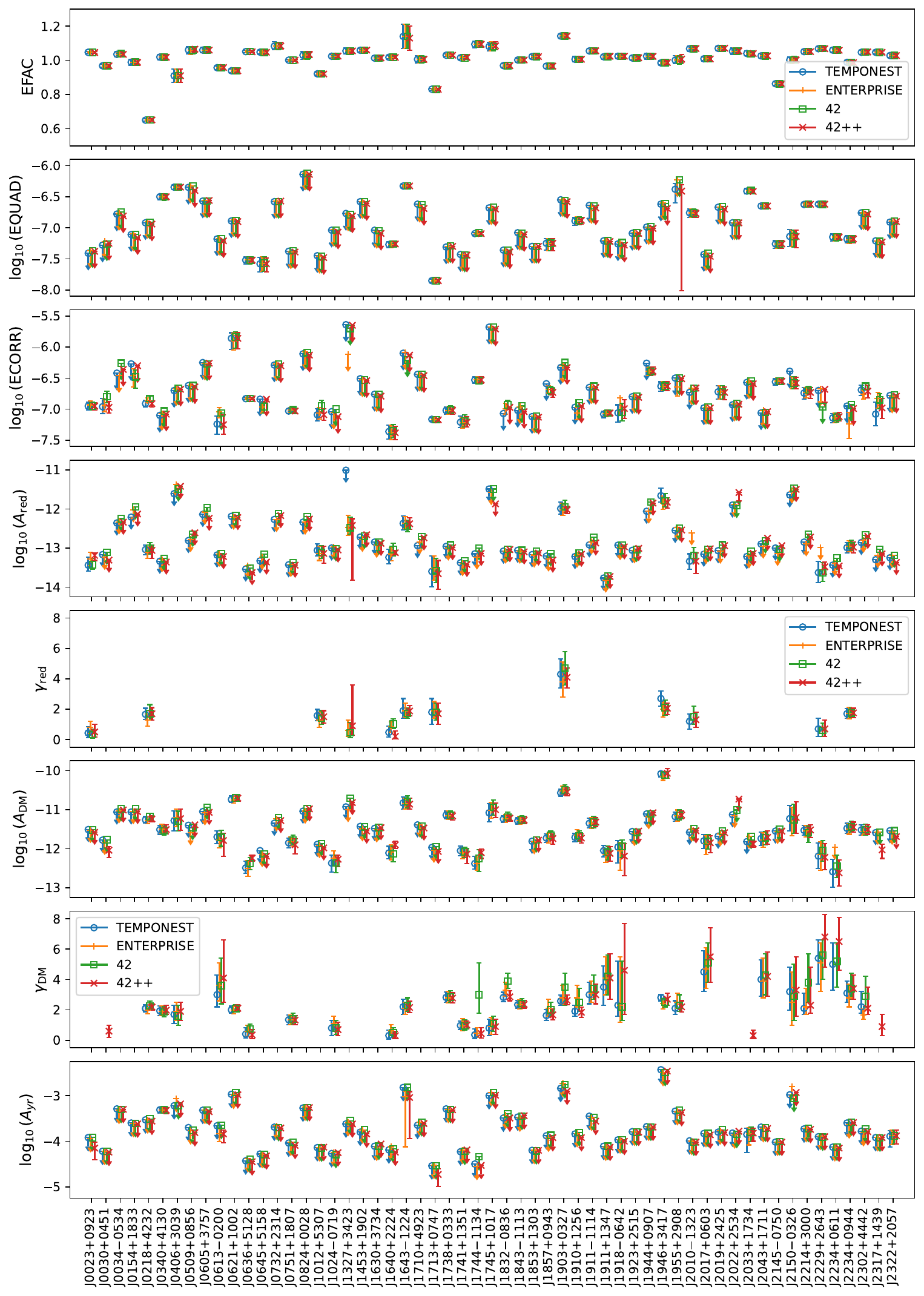}
\caption{Recovered noise parameters from the four different software packages. The median values and central 68\% credible regions are indicated by the points and error bars. The 95\% upper limits are shown with arrows pointing down.}
\label{fig:noise}
\end{figure*}

Our standard noise model for step 2 includes three white noise, two red noise, two DM GP, and two parameters for the annual DM variation. This step was performed using all four software packages. However, the Romberg weights were only applied in the \texttt{42++} analysis.
The tables in Appendix~\ref{sec:tab} show all these parameters for the four different pipelines.
The Bayesian sampling is done with all nine (five) parameters for the DM GP (DMX) models simultaneously.
Time series realizations of the red and DM GP noise can be found for two example pulsars in Figure~\ref{fig:gp_series}.
In general, all four pipelines give very consistent constraints or upper limits on the parameters, as demonstrated in Figure~\ref{fig:noise}.

As several parameters are poorly constrained by the CPTA DR1 with its approximately three years of observations, the inclusion of the corresponding noise terms could increase the computational cost unnecessarily and possibly decrease the sensitivity of the GW signal search. Therefore, we performed another round of model comparisons investigating EQUAD, ECORR, DM and red noise to determine which of the noise models is the most favoured. The analysis was done simultaneously with fitting timing model parameters. Step 3 was run with \texttt{TEMPONEST} only. The most favoured combination of noise terms can be found in Table 2 of Xu et al., in prep.

\subsection{DM GP versus DMX}

\begin{figure*}
\centering
\includegraphics[width=\linewidth]{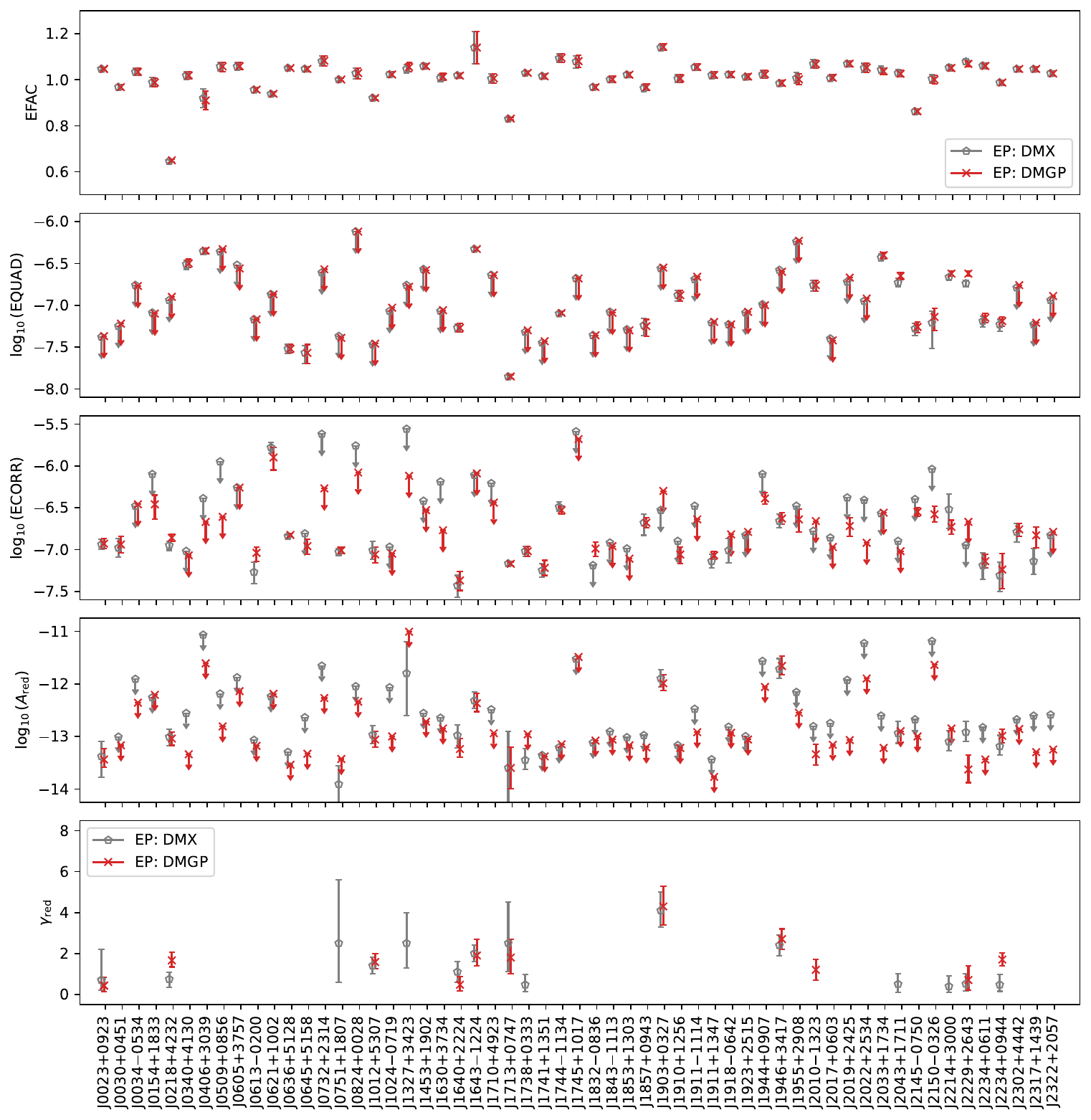}
\caption{Comparison between the white and red noise parameter constraints from the DM GP and DMX model with \texttt{ENTERPRISE}. The median values and central 68\% credible regions are indicated by the points and error bars. The 95\% upper limits are shown with arrows pointing down.}
\label{fig:dmgp_dmx}
\end{figure*}

We also performed steps 1 and 2 of the analysis for the dataset obtained with the DMX model. Figure~\ref{fig:dmgp_dmx} shows a comparison of the median and central 68\% for the white and red noise parameter posteriors between the DMX and DM GP models.

The recovered EFAC and EQUAD white noise parameters are identical for all pulsars. There are, however, noticeable differences in the posteriors of the ECORR parameter. This could be due to power at high frequencies that the DMX model can fit, while the power-law model has a high frequency cutoff. On the other hand, the DMX model could also be overfitting and absorb high frequency noise, which could lead to lower Bayesian evidence.

Most pulsars show broadly consistent red noise constraints. However, a general trend of larger upper limits in the DMX model can be seen. It is possible that some of the pulsar intrinsic red noise has been absorbed by the DMX model or that high frequency white noise is wrongly attributed to red noise by the DM GP model.

We would also like to point out that in all cases the Bayesian evidence for the DMX dataset is significantly lower compared to the DM GP model. We note that for most of our pulsars the logarithmic Bayesian evidence difference $\Delta \ln\mathcal{B}$ comparing the DM GP to the DMX model reaches the level of a few thousand. Thus, we focus on the analysis with the DM GP model \footnote{As a caveat, the direct comparison between the Bayesian evidence of the DMX and DM GP models may not be rigorous. The two models are not nested and the model parameters enter the DMX and DM GP model very differently, through fitting waveform (DMX) and covariance matrix (DM GP), respectively.}.

\subsection{Parameter estimation}

We summarize the noise properties of the 57 pulsars of the CPTA DR1. They are classified into four categories: 1. white noise only, 2. DM and white noise, 3. red and white noise, and 4. DM, red and white noise (see \TAB{tab:par}).

The EFAC white noise parameters are in general close to the expected value of 1, indicating that the initial estimate does not need to be modified much. Twenty pulsars have neither EQUAD nor ECORR constraints, thus we can only set upper limits for them (see \TAB{tab:wn_par}). Out of the remaining 37 pulsars, 19 (28) have constraints on EQUAD (ECORR) respectively, with 10 having both. The appearance of the ECORR parameter is expected as the observations are channelized, i.e. several TOAs are created from the same observation by splitting the full radio frequency bandwidth into 16 or 64 channels.

About 14 pulsars have constrained parameter values for the red noise power law and 32 pulsars have constrained values for the DM GP. For the other pulsars we set upper limits, however, there are 15 (9) additional pulsars that have Bayesian evidence supporting the inclusion of the red noise (DM variations) in the model.

Five pulsars are found to favour only white noise models: J0509+0856, J0605+3757, J0732+2314, J1710+4923, and J2019+2425, where the first two only require EFAC and the latter three have EFAC and ECORR. For 23 pulsars we found significant evidence for a DM variation modelled with a power-law GP in addition to the white noise, but no evidence for pulsar red noise. Evidence to include red noise on top of white noise was found in 11 pulsars. Eighteen pulsars have the highest Bayes factor for the full noise model with white, red, and DM noise all being measured.

All four software packages produce consistent results for almost all pulsars, with a few notable exceptions. J1327+3423 has a bimodal posterior distribution for the ECORR parameter. While \texttt{ENTERPRISE} only finds the lower of the two peaks, the other three software packages produce a bimodal distribution, thus producing a larger upper limit. This could be either due to the differences in the implementation or the different sampling methods. Other examples include J1643$-$1224, J2010$-$1323, and J2033+1734, where small differences in the tails of the distributions lead to different upper limits. Additionally, some pulsars, like J1832$-$0836, J1903+0327, and J1910+1256, have constrained values in some software packages, while the other software packages produce upper limits.

\setlength{\tabcolsep}{3pt}
\begin{table}
\renewcommand{\arraystretch}{1.3}
\caption{Number of pulsars with constrained parameters for noise terms in the model.}
\centering
\begin{tabular}{c|c|c|c}
\hline\hline
WN only & DM+WN & RN+WN & DM+RN+WN \\
\hline
20 (5)  & 23 (23) & 5 (11) & 9 (18) \\
\hline\hline
\end{tabular}
\label{tab:par}
\tablefoot{The numbers in brackets denote the number of pulsars with Bayesian evidence for certain noise terms although their parameters may not be fully constrained, see Xu et al., in prep.}
\end{table}

\setlength{\tabcolsep}{3pt}
\begin{table}
\renewcommand{\arraystretch}{1.3}
\caption{Number of pulsars with constrained white noise parameters.}
\centering
\begin{tabular}{c|c|c|c}
\hline\hline
EFAC only & EQUAD & ECORR & EQUAD+ECORR \\
\hline
20        & 9     & 18    & 10 \\
\hline\hline
\end{tabular}
\label{tab:wn_par}
\end{table}

\setlength{\tabcolsep}{3pt}
\begin{table}
\renewcommand{\arraystretch}{1.3}
\caption{Number of pulsars with constrained noise terms and their covariances.}
\centering
\begin{tabular}{ccc|cccc}
\hline\hline
\multicolumn{3}{c|}{Covariance} & \multicolumn{4}{c}{No covariance} \\ \hline
\multicolumn{3}{c|}{24}         & \multicolumn{4}{c}{33}            \\ \hline
\multicolumn{1}{c|}{DM-WN}      & \multicolumn{1}{c|}{RN-WN}        & \multicolumn{1}{c|}{DM-RN} & \multicolumn{3}{c|}{Constrained} & Unconstrained \\ \hline
\multicolumn{1}{c|}{4}          & \multicolumn{1}{c|}{10}           & \multicolumn{1}{c|}{5}     & \multicolumn{3}{c|}{26}          & 7             \\ \cline{1-3} \cline{4-6}
\multicolumn{3}{c|}{DM-RN-WN}   & \multicolumn{1}{c|}{DM}           & \multicolumn{1}{c|}{RN}    & \multicolumn{1}{c|}{DM+RN}       & \\ \cline{1-3} \cline{4-6}
\multicolumn{3}{c|}{5}          & \multicolumn{1}{c|}{19}           & \multicolumn{1}{c|}{3}     & \multicolumn{1}{c|}{4}           & \\
\hline\hline
\end{tabular}
\label{tab:cor}
\end{table}

\subsection{Parameter covariances}

Next, we investigate the covariances between different noise terms and parameters. The covariance between red noise and DM variations is expected as both cause long-term trends in the residuals. This covariance can be broken with multiple radio frequency coverage and longer observation time spans. For the amplitude parameter, posterior covariances usually result in distributions that have a peak, but also a long tail, resulting in the typical 'L'-shape in the 2D posterior distributions. \TAB{tab:cor} summarizes the number of pulsars in the different categories detailed below. The full 2D posteriors can be found in the online supplementary material \footnote{\url{https://doi.org/10.5281/zenodo.13828113}}.

\subsubsection{DM GP -- DM annual}
For the CPTA DR1 with three years of observations, the annual DM signal is highly covariate with the DM GP power-law model. Although we only measure constrained values for the annual DM amplitude $A_{\rm yr}$ for about three pulsars (J0340+4130, J2033+1734, and J2322+2057), five more show an indication of a measurement (J0023+0923, J0824+0028, J1012+5307, J2234+0611, and J2234+0944), and 12 show a strong correlation between the DM GP power law and annual amplitude (J0154+1833, J0613$-$0200, J1643$-$1224, J1713+0747, J1738+0333, J1911+1347, J1918$-$0642, J1944+0907, J2017+0603, J2033+1734, J2145$-$0750, and J2317+1439). Longer observation time spans will help to break degeneracies.

\subsubsection{Red noise -- DM}
The expected correlation between pulsar red noise and DM GP variations have been found in ten pulsars. However, five of these also show covariance with the ECORR white noise parameter (see next section). The five pulsars with purely red and DM noise covariance are J0030+0451, J0034$-$0534, J1853+1303, J1918$-$0642, and J1923+2515.

\subsubsection{Red noise -- DM -- white noise}
In addition to the covariance between red and DM noise, we also measured some degree of interplay between `long'-term variations and `short'-term white noise. Given that the time span is only three years, the distinction between red, DM, and white noise can be difficult. In total, 19 pulsars show difficulties in distinguishing the three noise terms. Five of these pulsars have a strong covariance between the ECORR parameter and both red and DM noise: J0154+1833, J0824+0028, J1903+0327, J2317+1439, and J2322+2057. The remaining 14 pulsars either show covariance between red and white noise (four pulsars) or DM and white noise (ten pulsars).

\subsubsection{Red noise -- white noise}
As a sub-class of the red-DM-white noise covariance, ten pulsars only show interplay between the red and white noise. The DM parameters are well constrained for four pulsars (J0613$-$0200, J0621+1002, J1643$-$1224, and J2229+2643), while they are set as upper limits for the other six pulsars (J0645+5158, J1327+3423, J1453+1902, J1630+3734, J2010$-$1323, and J2022+2534).

\subsubsection{DM -- white noise}
Another sub-class of the red-DM-white noise covariance are the DM-white noise pulsars, which are the opposite of red-white noise case. Only four pulsars show this type of interplay: J1910+1256, J1955+2908, J2033+1734, and J2214+3000. In all cases, the red noise term is not constrained.

\subsubsection{No covariances}
Thirty-three pulsars in the CPTA DR1 show no covariances between the noise terms. This is the case for the 26 pulsars that have well-constrained parameter distributions for red and/or DM noise, as well as seven pulsars that show no sign of red or DM noise terms. The seven pulsars with white noise only are J0509+0856, J0605+3757, J0732+2314, J1710+4923, J1944+0907, J2019+2425, and J2145$-$0750.

\subsubsection{Covariances in the DMX dataset}
As a comparison, we briefly summarize the constrained parameters and their covariances in the DMX model with only white and red noise. The figures showing the full 2D posterior distributions can be found in the online supplementary material \footnote{\url{https://doi.org/10.5281/zenodo.13828113}}.
Two pulsars show interplay between the EQUAD and ECORR parameters: J1955+2908 and J2150$-$0326. The similarity between the two parameters could explain the covariance.
Eighteen pulsars have an indication for red noise, with ten of them with constrained parameters and eight of them sharing a covariance with the ECORR parameter.

\section{Discussion and conclusion}
\label{sec:discussion}

We compared the parameter constraints for the red noise and DM GP power laws with the EPTA DR2 \citep{eptadr2_noise}, NG15 \citep{ng15_noise}, and PPTA DR3 \citep{pptadr3_noise} for a total of 22 overlapping pulsars, see Figure~\ref{fig:comp}. As each PTA uses different settings to produce the final noise models, the figures should be viewed with these caveats in mind. The EPTA DR2 has two versions of the dataset and only includes a noise term if it is significant. The PPTA DR3 noise analysis includes other noise terms that we have not considered. Finally, NANOGrav primarily uses the DMX model, thus there are no DM GP constraints to compare against. One can see very broad agreement between the CPTA results and those from the other PTAs. \cite{ipta_comp} provide a more detailed comparison between the noise models from the different PTAs.

The CPTA DR1 obtained using about three years of FAST observations is an unprecedentedly sensitive PTA dataset for GW searches. As such, we have investigated in detail the timing and noise analysis of each of the 57 pulsars in the dataset. The results for the timing model parameters and other details of the dataset can be found in Xu et al., in prep. Here, the results of the noise models have been presented. We followed a three-step process to determine the optimal frequency bin choice for the power-law model, verify the parameter posterior distributions with four different software packages, and finally use Bayesian analysis to select the optimal noise terms to be included in the final timing and noise model.

The analysis was performed for both DM GP and DMX style datasets. The DMX model allows for larger red noise amplitudes for the majority of the pulsars and has significantly lower Bayesian evidence compared to the DM GP model. The DM GP dataset has been chosen as our main dataset. In general, the four software packages, \texttt{TEMPONEST}, \texttt{ENTERPRISE}, \texttt{42,} and \texttt{42++}, produce very consistent constraints on all the parameters in the noise model.

For the majority of pulsars, we do not measure red or DM noise and therefore set upper limits on the amplitudes. Fourteen (32) pulsars show constraints for the red (DM) noise parameters, while an additional 15 (9) have Bayesian evidence for the respective noise term to be included (Xu et al., in prep.). We also tested the conventional description for the EQUAD and ECORR parameters against a time-dependent jitter noise description and found no conclusive evidence for either model. The hypothesis that ECORR should be linked to the pulsar jitter process is further tested in Wang et al., in prep.

We find no covariances between the parameters of the different noise terms for 33 pulsars, the majority of which have constrained posteriors for the parameters. The remaining 24 pulsars display covariances between red, DM, and white noise, notably ECORR. Given the short length of the dataset, it may be difficult to clearly distinguish between long-term red/DM variations and short-term white noise. While in this paper we focus on the noise terms with constrained parameters and their covariances, the final timing and noise models used for the GW search \citep{cpta_dr1} were obtained from Bayesian model comparison in Xu et al, in prep.

The CPTA DR1 is opening a new frequency range that has not been accessible by the established PTA collaborations, with much increased sensitivity and reduced uncertainty using FAST observations. Being only about three years long, long-term variations cannot yet be well probed with the CPTA data; continued observation of the 57 pulsars will allow us to break the degeneracies found between different noise terms with future CPTA datasets. We plan to join the efforts in the IPTA to combine datasets from all PTA collaborations in the future. Such an IPTA dataset will significantly improve the GW science that can be done with a PTA.

\begin{acknowledgements}

Observation of CPTA is supported by the FAST Key project. This work has used the data from the Five-hundred-meter Aperture Spherical radio Telescope (FAST, https://cstr.cn/31116.02.FAST). FAST is a Chinese national mega-science facility, operated by National Astronomical Observatories, Chinese Academy of Sciences (NAOC). This work is supported by the National SKA Program of China (2020SKA0120100), the National Key Research and Development Program of China No.2022YFC2205203, the National Natural Science Foundation of China grant no. 12041303, 12250410246, 12173087 and 12063003, China Postdoctoral Science Foundation No. 2023M743518 and 2023M743516, Major Science and Technology Program of Xinjiang Uygur Autonomous Region No. 2022A03013-4, the CAS-MPG LEGACY project, and funding from the Max-Planck Partner Group. KJL acknowledges support from the XPLORER PRIZE and 20-year long-term support from Dr. Guojun Qiao. The data analysis are performed with computer clusters \textsc{DIRAC} and \textsc{C*-system} of PSR@pku and computational resource provide by the \textsc{PARATERA} company. Software package 42++ is developed with \textsc{Intel} oneAPI toolkits and the \textsc{Science Explorer's Developing GEars} from \textsc{Weichuan technology}. We thank Dr. Jin Chang for providing the valuable long-term support to the CPTA collaboration. CPTA thanks Dr Duncan Lorimer, Dr Michael Kramer and Dr Richard Manchester for helping initialize the collaboration. SC would like to thank Dr Bruce Allen for his comments on the description of the analysis methods.

\end{acknowledgements}

\bibliographystyle{aa}
\bibliography{bibliography}

\begin{thebibliography}{43}
\expandafter\ifx\csname natexlab\endcsname\relax\def\natexlab#1{#1}\fi

\bibitem[{{Agazie} {et~al.}(2024){Agazie}, {Antoniadis}, {Anumarlapudi}, {Archibald}, {Arumugam}, {Arumugam}, {Arzoumanian}, {Askew}, {Babak}, {Bagchi}, {Bailes}, {Bak Nielsen}, {Baker}, {Bassa}, {Bathula}, {B{\'e}csy}, {Berthereau}, {Bhat}, {Blecha}, {Bonetti}, {Bortolas}, {Brazier}, {Brook}, {Burgay}, {Burke-Spolaor}, {Burnette}, {Caballero}, {Cameron}, {Case}, {Chalumeau}, {Champion}, {Chanlaridis}, {Charisi}, {Chatterjee}, {Chatziioannou}, {Cheeseboro}, {Chen}, {Chen}, {Cognard}, {Cohen}, {Coles}, {Cordes}, {Cornish}, {Crawford}, {Cromartie}, {Crowter}, {Cury{\l}o}, {Cutler}, {Dai}, {Dandapat}, {Deb}, {DeCesar}, {DeGan}, {Demorest}, {Deng}, {Desai}, {Desvignes}, {Dey}, {Dhanda-Batra}, {Di Marco}, {Dolch}, {Drachler}, {Dwivedi}, {Ellis}, {Falxa}, {Feng}, {Ferdman}, {Ferrara}, {Fiore}, {Fonseca}, {Franchini}, {Freedman}, {Gair}, {Garver-Daniels}, {Gentile}, {Gersbach}, {Glaser}, {Good}, {Goncharov}, {Gopakumar}, {Graikou}, {Griessmeier}, {Guillemot}, {G{\"u}ltekin}, {Guo}, {Gupta}, {Grunthal}, {Hazboun},
  {Hisano}, {Hobbs}, {Hourihane}, {Hu}, {Iraci}, {Islo}, {Izquierdo-Villalba}, {Jang}, {Jawor}, {Janssen}, {Jennings}, {Jessner}, {Johnson}, {Jones}, {Joshi}, {Kaiser}, {Kaplan}, {Kapur}, {Kareem}, {Karuppusamy}, {Keane}, {Keith}, {Kelley}, {Kerr}, {Key}, {Kharbanda}, {Kikunaga}, {Klein}, {Kolhe}, {Kramer}, {Krishnakumar}, {Kulkarni}, {Laal}, {Lackeos}, {Lam}, {Lamb}, {Larsen}, {Lazio}, {Lee}, {Levin}, {Lewandowska}, {Littenberg}, {Liu}, {Liu}, {Liu}, {Lommen}, {Lorimer}, {Lower}, {Luo}, {Luo}, {Lynch}, {Lyne}, {Ma}, {Maan}, {Madison}, {Main}, {Manchester}, {Mandow}, {Mattson}, {McEwen}, {McKee}, {McLaughlin}, {McMann}, {Meyers}, {Meyers}, {Mickaliger}, {Miles}, {Mingarelli}, {Mitridate}, {Natarajan}, {Nathan}, {Ng}, {Nice}, {Ni{\c{t}}u}, {Nobleson}, {Ocker}, {Olum}, {Os{\l}owski}, {Paladi}, {Parthasarathy}, {Pennucci}, {Perera}, {Perrodin}, {Petiteau}, {Petrov}, {Pol}, {Porayko}, {Possenti}, {Prabu}, {Quelquejay Leclere}, {Radovan}, {Rana}, {Ransom}, {Ray}, {Reardon}, {Rogers}, {Romano}, {Russell},
  {Samajdar}, {Sanidas}, {Sardesai}, {Schmiedekamp}, {Schmiedekamp}, {Schmitz}, {Schult}, {Sesana}, {Shaifullah}, {Shannon}, {Shapiro-Albert}, {Siemens}, {Simon}, {Singha}, {Siwek}, {Speri}, {Spiewak}, {Srivastava}, {Stairs}, {Stappers}, {Stinebring}, {Stovall}, {Sun}, {Surnis}, {Susarla}, {Susobhanan}, {Swiggum}, {Takahashi}, {Tarafdar}, {Taylor}, {Taylor}, {Theureau}, {Thrane}, {Thyagarajan}, {Tiburzi}, {Toomey}, {Turner}, {Unal}, {Vallisneri}, {van der Wateren}, {van Haasteren}, {Vecchio}, {Venkatraman Krishnan}, {Verbiest}, {Vigeland}, {Wahl}, {Wang}, {Wang}, {Witt}, {Wang}, {Wang}, {Wayt}, {Wu}, {Young}, {Zhang}, {Zhang}, {Zhu}, {Zic}, \& {International Pulsar Timing Array Collaboration}}]{ipta_comp}
{Agazie}, G., {Antoniadis}, J., {Anumarlapudi}, A., {et~al.} 2024, \apj, 966, 105

\bibitem[{{Agazie} {et~al.}(2023{\natexlab{a}}){Agazie}, {Anumarlapudi}, {Archibald}, {Arzoumanian}, {Baker}, {B{\'e}csy}, {Blecha}, {Brazier}, {Brook}, {Burke-Spolaor}, {Burnette}, {Case}, {Charisi}, {Chatterjee}, {Chatziioannou}, {Cheeseboro}, {Chen}, {Cohen}, {Cordes}, {Cornish}, {Crawford}, {Cromartie}, {Crowter}, {Cutler}, {Decesar}, {Degan}, {Demorest}, {Deng}, {Dolch}, {Drachler}, {Ellis}, {Ferrara}, {Fiore}, {Fonseca}, {Freedman}, {Garver-Daniels}, {Gentile}, {Gersbach}, {Glaser}, {Good}, {G{\"u}ltekin}, {Hazboun}, {Hourihane}, {Islo}, {Jennings}, {Johnson}, {Jones}, {Kaiser}, {Kaplan}, {Kelley}, {Kerr}, {Key}, {Klein}, {Laal}, {Lam}, {Lamb}, {Lazio}, {Lewandowska}, {Littenberg}, {Liu}, {Lommen}, {Lorimer}, {Luo}, {Lynch}, {Ma}, {Madison}, {Mattson}, {McEwen}, {McKee}, {McLaughlin}, {McMann}, {Meyers}, {Meyers}, {Mingarelli}, {Mitridate}, {Natarajan}, {Ng}, {Nice}, {Ocker}, {Olum}, {Pennucci}, {Perera}, {Petrov}, {Pol}, {Radovan}, {Ransom}, {Ray}, {Romano}, {Sardesai}, {Schmiedekamp}, {Schmiedekamp},
  {Schmitz}, {Schult}, {Shapiro-Albert}, {Siemens}, {Simon}, {Siwek}, {Stairs}, {Stinebring}, {Stovall}, {Sun}, {Susobhanan}, {Swiggum}, {Taylor}, {Taylor}, {Turner}, {Unal}, {Vallisneri}, {van Haasteren}, {Vigeland}, {Wahl}, {Wang}, {Witt}, {Young}, \& {Nanograv Collaboration}}]{ng15_gwb}
{Agazie}, G., {Anumarlapudi}, A., {Archibald}, A.~M., {et~al.} 2023{\natexlab{a}}, \apjl, 951, L8

\bibitem[{{Agazie} {et~al.}(2023{\natexlab{b}}){Agazie}, {Anumarlapudi}, {Archibald}, {Arzoumanian}, {Baker}, {B{\'e}csy}, {Blecha}, {Brazier}, {Brook}, {Burke-Spolaor}, {Charisi}, {Chatterjee}, {Cohen}, {Cordes}, {Cornish}, {Crawford}, {Cromartie}, {Crowter}, {Decesar}, {Demorest}, {Dolch}, {Drachler}, {Ferrara}, {Fiore}, {Fonseca}, {Freedman}, {Garver-Daniels}, {Gentile}, {Glaser}, {Good}, {Guertin}, {G{\"u}ltekin}, {Hazboun}, {Jennings}, {Johnson}, {Jones}, {Kaiser}, {Kaplan}, {Kelley}, {Kerr}, {Key}, {Laal}, {Lam}, {Lamb}, {Lazio}, {Lewandowska}, {Liu}, {Lorimer}, {Luo}, {Lynch}, {Ma}, {Madison}, {McEwen}, {McKee}, {McLaughlin}, {McMann}, {Meyers}, {Mingarelli}, {Mitridate}, {Ng}, {Nice}, {Ocker}, {Olum}, {Pennucci}, {Perera}, {Pol}, {Radovan}, {Ransom}, {Ray}, {Romano}, {Sardesai}, {Schmiedekamp}, {Schmiedekamp}, {Schmitz}, {Shapiro-Albert}, {Siemens}, {Simon}, {Siwek}, {Stairs}, {Stinebring}, {Stovall}, {Susobhanan}, {Swiggum}, {Taylor}, {Turner}, {Unal}, {Vallisneri}, {Vigeland}, {Wahl}, {Witt},
  {Young}, \& {Nanograv Collaboration}}]{ng15_noise}
{Agazie}, G., {Anumarlapudi}, A., {Archibald}, A.~M., {et~al.} 2023{\natexlab{b}}, \apjl, 951, L10

\bibitem[{{Arzoumanian} {et~al.}(2016){Arzoumanian}, {Brazier}, {Burke-Spolaor}, {Chamberlin}, {Chatterjee}, {Christy}, {Cordes}, {Cornish}, {Crowter}, {Demorest}, {Deng}, {Dolch}, {Ellis}, {Ferdman}, {Fonseca}, {Garver-Daniels}, {Gonzalez}, {Jenet}, {Jones}, {Jones}, {Kaspi}, {Koop}, {Lam}, {Lazio}, {Levin}, {Lommen}, {Lorimer}, {Luo}, {Lynch}, {Madison}, {McLaughlin}, {McWilliams}, {Mingarelli}, {Nice}, {Palliyaguru}, {Pennucci}, {Ransom}, {Sampson}, {Sanidas}, {Sesana}, {Siemens}, {Simon}, {Stairs}, {Stinebring}, {Stovall}, {Swiggum}, {Taylor}, {Vallisneri}, {van Haasteren}, {Wang}, {Zhu}, \& {NANOGrav Collaboration}}]{ng9}
{Arzoumanian}, Z., {Brazier}, A., {Burke-Spolaor}, S., {et~al.} 2016, \apj, 821, 13

\bibitem[{{Caballero} {et~al.}(2016){Caballero}, {Lee}, {Lentati}, {Desvignes}, {Champion}, {Verbiest}, {Janssen}, {Stappers}, {Kramer}, {Lazarus}, {Possenti}, {Tiburzi}, {Perrodin}, {Os{\l}owski}, {Babak}, {Bassa}, {Brem}, {Burgay}, {Cognard}, {Gair}, {Graikou}, {Guillemot}, {Hessels}, {Karuppusamy}, {Lassus}, {Liu}, {McKee}, {Mingarelli}, {Petiteau}, {Purver}, {Rosado}, {Sanidas}, {Sesana}, {Shaifullah}, {Smits}, {Taylor}, {Theureau}, {van Haasteren}, \& {Vecchio}}]{cll+2016}
{Caballero}, R.~N., {Lee}, K.~J., {Lentati}, L., {et~al.} 2016, \mnras, 457, 4421

\bibitem[{{Chalumeau} {et~al.}(2022){Chalumeau}, {Babak}, {Petiteau}, {Chen}, {Samajdar}, {Caballero}, {Theureau}, {Guillemot}, {Desvignes}, {Parthasarathy}, {Liu}, {Shaifullah}, {Hu}, {van der Wateren}, {Antoniadis}, {Bak Nielsen}, {Bassa}, {Berthereau}, {Burgay}, {Champion}, {Cognard}, {Falxa}, {Ferdman}, {Freire}, {Gair}, {Graikou}, {Guo}, {Jang}, {Janssen}, {Karuppusamy}, {Keith}, {Kramer}, {Lee}, {Liu}, {Lyne}, {Main}, {McKee}, {Mickaliger}, {Perera}, {Perrodin}, {Porayko}, {Possenti}, {Sanidas}, {Sesana}, {Speri}, {Stappers}, {Tiburzi}, {Vecchio}, {Verbiest}, {Wang}, {Wang}, \& {Xu}}]{cbp+2022}
{Chalumeau}, A., {Babak}, S., {Petiteau}, A., {et~al.} 2022, \mnras, 509, 5538

\bibitem[{Chapra \& Canale(2014)}]{NME}
Chapra, S.~C. \& Canale, R.~P. 2014, {Numerical Methods for Engineers} ({McGraw-Hill Education, 2 Penn Plaza, New York, NY 10121})

\bibitem[{{Chen} {et~al.}(2021){Chen}, {Caballero}, {Guo}, {Chalumeau}, {Liu}, {Shaifullah}, {Lee}, {Babak}, {Desvignes}, {Parthasarathy}, {Hu}, {van der Wateren}, {Antoniadis}, {Bak Nielsen}, {Bassa}, {Berthereau}, {Burgay}, {Champion}, {Cognard}, {Falxa}, {Ferdman}, {Freire}, {Gair}, {Graikou}, {Guillemot}, {Jang}, {Janssen}, {Karuppusamy}, {Keith}, {Kramer}, {Liu}, {Lyne}, {Main}, {McKee}, {Mickaliger}, {Perera}, {Perrodin}, {Petiteau}, {Porayko}, {Possenti}, {Samajdar}, {Sanidas}, {Sesana}, {Speri}, {Stappers}, {Theureau}, {Tiburzi}, {Vecchio}, {Verbiest}, {Wang}, {Wang}, \& {Xu}}]{ccg+2021}
{Chen}, S., {Caballero}, R.~N., {Guo}, Y.~J., {et~al.} 2021, \mnras, 508, 4970

\bibitem[{Ellis \& van Haasteren(2017)}]{ptmcmc}
Ellis, J. \& van Haasteren, R. 2017, jellis18/PTMCMCSampler: Official Release

\bibitem[{Ellis {et~al.}(2020)Ellis, Vallisneri, Taylor, \& Baker}]{enterprise}
Ellis, J.~A., Vallisneri, M., Taylor, S.~R., \& Baker, P.~T. 2020, ENTERPRISE: Enhanced Numerical Toolbox Enabling a Robust PulsaR Inference SuitE, Zenodo

\bibitem[{{EPTA Collaboration} {et~al.}(2023{\natexlab{a}}){EPTA Collaboration}, {InPTA Collaboration}, {Antoniadis}, {Arumugam}, {Arumugam}, {Babak}, {Bagchi}, {Bak Nielsen}, {Bassa}, {Bathula}, {Berthereau}, {Bonetti}, {Bortolas}, {Brook}, {Burgay}, {Caballero}, {Chalumeau}, {Champion}, {Chanlaridis}, {Chen}, {Cognard}, {Dandapat}, {Deb}, {Desai}, {Desvignes}, {Dhanda-Batra}, {Dwivedi}, {Falxa}, {Ferdman}, {Franchini}, {Gair}, {Goncharov}, {Gopakumar}, {Graikou}, {Grie{\ss}meier}, {Guillemot}, {Guo}, {Gupta}, {Hisano}, {Hu}, {Iraci}, {Izquierdo-Villalba}, {Jang}, {Jawor}, {Janssen}, {Jessner}, {Joshi}, {Kareem}, {Karuppusamy}, {Keane}, {Keith}, {Kharbanda}, {Kikunaga}, {Kolhe}, {Kramer}, {Krishnakumar}, {Lackeos}, {Lee}, {Liu}, {Liu}, {Lyne}, {McKee}, {Maan}, {Main}, {Mickaliger}, {Ni{\c{t}}u}, {Nobleson}, {Paladi}, {Parthasarathy}, {Perera}, {Perrodin}, {Petiteau}, {Porayko}, {Possenti}, {Prabu}, {Quelquejay Leclere}, {Rana}, {Samajdar}, {Sanidas}, {Sesana}, {Shaifullah}, {Singha}, {Speri}, {Spiewak},
  {Srivastava}, {Stappers}, {Surnis}, {Susarla}, {Susobhanan}, {Takahashi}, {Tarafdar}, {Theureau}, {Tiburzi}, {van der Wateren}, {Vecchio}, {Venkatraman Krishnan}, {Verbiest}, {Wang}, {Wang}, \& {Wu}}]{eptadr2_gwb}
{EPTA Collaboration}, {InPTA Collaboration}, {Antoniadis}, J., {et~al.} 2023{\natexlab{a}}, \aap, 678, A50

\bibitem[{{EPTA Collaboration} {et~al.}(2023{\natexlab{b}}){EPTA Collaboration}, {InPTA Collaboration}, {Antoniadis}, {Arumugam}, {Arumugam}, {Babak}, {Bagchi}, {Nielsen}, {Bassa}, {Bathula}, {Berthereau}, {Bonetti}, {Bortolas}, {Brook}, {Burgay}, {Caballero}, {Chalumeau}, {Champion}, {Chanlaridis}, {Chen}, {Cognard}, {Dandapat}, {Deb}, {Desai}, {Desvignes}, {Dhanda-Batra}, {Dwivedi}, {Falxa}, {Ferdman}, {Franchini}, {Gair}, {Goncharov}, {Gopakumar}, {Graikou}, {Grie{\ss}meier}, {Guillemot}, {Guo}, {Gupta}, {Hisano}, {Hu}, {Iraci}, {Izquierdo-Villalba}, {Jang}, {Jawor}, {Janssen}, {Jessner}, {Joshi}, {Kareem}, {Karuppusamy}, {Keane}, {Keith}, {Kharbanda}, {Kikunaga}, {Kolhe}, {Kramer}, {Krishnakumar}, {Lackeos}, {Lee}, {Liu}, {Liu}, {Lyne}, {McKee}, {Maan}, {Main}, {Mickaliger}, {Ni{\c{t}}u}, {Nobleson}, {Paladi}, {Parthasarathy}, {Perera}, {Perrodin}, {Petiteau}, {Porayko}, {Possenti}, {Prabu}, {Leclere}, {Rana}, {Samajdar}, {Sanidas}, {Sesana}, {Shaifullah}, {Singha}, {Speri}, {Spiewak}, {Srivastava},
  {Stappers}, {Surnis}, {Susarla}, {Susobhanan}, {Takahashi}, {Tarafdar}, {Theureau}, {Tiburzi}, {van der Wateren}, {Vecchio}, {Krishnan}, {Verbiest}, {Wang}, {Wang}, \& {Wu}}]{eptadr2_noise}
{EPTA Collaboration}, {InPTA Collaboration}, {Antoniadis}, J., {et~al.} 2023{\natexlab{b}}, \aap, 678, A49

\bibitem[{Falxa {et~al.}(2023)Falxa, Babak, Baker, Bécsy, Chalumeau, Chen, Chen, Cornish, Guillemot, Hazboun, Mingarelli, Parthasarathy, Petiteau, Pol, Sesana, Spolaor, Taylor, Theureau, Vallisneri, Vigeland, Witt, Zhu, Antoniadis, Arzoumanian, Bailes, Bhat, Blecha, Brazier, Brook, Caballero, Cameron, Casey-Clyde, Champion, Charisi, Chatterjee, Cognard, Cordes, Crawford, Cromartie, Crowter, Dai, DeCesar, Demorest, Desvignes, Dolch, Drachler, Feng, Ferrara, Fiore, Fonseca, Garver-Daniels, Glaser, Goncharov, Good, Griessmeier, Guo, Gültekin, Hobbs, Hu, Islo, Jang, Jennings, Johnson, Jones, Kaczmarek, Kaiser, Kaplan, Keith, Kelley, Kerr, Key, Laal, Lam, Lamb, Lazio, Liu, Liu, Luo, Lynch, Madison, Main, Manchester, McEwen, McKee, McLaughlin, Ng, Nice, Ocker, Olum, Osłowski, Pennucci, Perera, Perrodin, Porayko, Possenti, Quelquejay-Leclere, Ransom, Ray, Reardon, Russell, Samajdar, Sarkissian, Schult, Shaifullah, Shannon, Shapiro-Albert, Siemens, Simon, Siwek, Smith, Speri, Spiewak, Stairs, Stappers, Stinebring,
  Swiggum, Tiburzi, Turner, Vecchio, Verbiest, Wahl, Wang, Wang, Wang, Wu, Zhang, Zhang, \& Collaboration}]{ipta_dr2_cgw}
Falxa, M., Babak, S., Baker, P.~T., {et~al.} 2023, Monthly Notices of the Royal Astronomical Society, 521, 5077

\bibitem[{{Feroz} \& {Hobson}(2008)}]{fh2008}
{Feroz}, F. \& {Hobson}, M.~P. 2008, \mnras, 384, 449

\bibitem[{{Foster} \& {Backer}(1990)}]{fb1990}
{Foster}, R.~S. \& {Backer}, D.~C. 1990, \apj, 361, 300

\bibitem[{{Goncharov} {et~al.}(2021){Goncharov}, {Reardon}, {Shannon}, {Zhu}, {Thrane}, {Bailes}, {Bhat}, {Dai}, {Hobbs}, {Kerr}, {Manchester}, {Os{\l}owski}, {Parthasarathy}, {Russell}, {Spiewak}, {Thyagarajan}, \& {Wang}}]{grs+2021}
{Goncharov}, B., {Reardon}, D.~J., {Shannon}, R.~M., {et~al.} 2021, \mnras, 502, 478

\bibitem[{{Hellings} \& {Downs}(1983)}]{hd1983}
{Hellings}, R.~W. \& {Downs}, G.~S. 1983, \apjl, 265, L39

\bibitem[{{Hobbs} {et~al.}(2006){Hobbs}, {Edwards}, \& {Manchester}}]{tempo2}
{Hobbs}, G.~B., {Edwards}, R.~T., \& {Manchester}, R.~N. 2006, \mnras, 369, 655

\bibitem[{{Jiang} {et~al.}(2019){Jiang}, {Yue}, {Gan}, {Yao}, {Li}, {Pan}, {Sun}, {Yu}, {Liu}, {Tang}, {Qian}, {Lu}, {Yan}, {Peng}, {Zhang}, {Wang}, {Li}, \& {Li}}]{jyg+2019}
{Jiang}, P., {Yue}, Y., {Gan}, H., {et~al.} 2019, Science China Physics, Mechanics, and Astronomy, 62, 959502

\bibitem[{Kass \& Raftery(1995)}]{kr1995}
Kass, R.~E. \& Raftery, A.~E. 1995, J. Amer. Stat. Assoc., 90, 773

\bibitem[{{Kramer} \& {Champion}(2013)}]{epta2013}
{Kramer}, M. \& {Champion}, D.~J. 2013, Classical and Quantum Gravity, 30, 224009

\bibitem[{{Larsen} {et~al.}(2024){Larsen}, {Mingarelli}, {Hazboun}, {Chalumeau}, {Good}, {Simon}, {Agazie}, {Anumarlapudi}, {Archibald}, {Arzoumanian}, {Baker}, {Brook}, {Cromartie}, {Crowter}, {DeCesar}, {Demorest}, {Dolch}, {Ferrara}, {Fiore}, {Fonseca}, {Freedman}, {Garver-Daniels}, {Gentile}, {Glaser}, {Jennings}, {Jones}, {Kaplan}, {Kerr}, {Lam}, {Lorimer}, {Luo}, {Lynch}, {McEwen}, {McLaughlin}, {McMann}, {Meyers}, {Ng}, {Nice}, {Pennucci}, {Perera}, {Pol}, {Radovan}, {Ransom}, {Ray}, {Schmiedekamp}, {Schmiedekamp}, {Shapiro-Albert}, {Stairs}, {Stovall}, {Susobhanan}, {Swiggum}, {Wahl}, {Champion}, {Cognard}, {Guillemot}, {Hu}, {Keith}, {Liu}, {McKee}, {Parthasarathy}, {Perrodin}, {Possenti}, {Shaifullah}, \& {Theureau}}]{lmh+2024}
{Larsen}, B., {Mingarelli}, C. M.~F., {Hazboun}, J.~S., {et~al.} 2024, \apj, 972, 49

\bibitem[{{Lee}(2016)}]{lee2016}
{Lee}, K.~J. 2016, in Astronomical Society of the Pacific Conference Series, Vol. 502, Frontiers in Radio Astronomy and FAST Early Sciences Symposium 2015, ed. L.~{Qain} \& D.~{Li}, 19

\bibitem[{{Lentati} {et~al.}(2014){Lentati}, {Alexander}, {Hobson}, {Feroz}, {van Haasteren}, {Lee}, \& {Shannon}}]{lah+2014}
{Lentati}, L., {Alexander}, P., {Hobson}, M.~P., {et~al.} 2014, \mnras, 437, 3004

\bibitem[{{Manchester} {et~al.}(2013){Manchester}, {Hobbs}, {Bailes}, {Coles}, {van Straten}, {Keith}, {Shannon}, {Bhat}, {Brown}, {Burke-Spolaor}, {Champion}, {Chaudhary}, {Edwards}, {Hampson}, {Hotan}, {Jameson}, {Jenet}, {Kesteven}, {Khoo}, {Kocz}, {Maciesiak}, {Oslowski}, {Ravi}, {Reynolds}, {Sarkissian}, {Verbiest}, {Wen}, {Wilson}, {Yardley}, {Yan}, \& {You}}]{ppta2013}
{Manchester}, R.~N., {Hobbs}, G., {Bailes}, M., {et~al.} 2013, \pasa, 30, e017

\bibitem[{{McLaughlin}(2013)}]{nanograv2013}
{McLaughlin}, M.~A. 2013, Classical and Quantum Gravity, 30, 224008

\bibitem[{{Miles} {et~al.}(2023){Miles}, {Shannon}, {Bailes}, {Reardon}, {Keith}, {Cameron}, {Parthasarathy}, {Shamohammadi}, {Spiewak}, {van Straten}, {Buchner}, {Camilo}, {Geyer}, {Karastergiou}, {Kramer}, {Serylak}, {Theureau}, \& {Venkatraman Krishnan}}]{mpta_dr1}
{Miles}, M.~T., {Shannon}, R.~M., {Bailes}, M., {et~al.} 2023, \mnras, 519, 3976

\bibitem[{{Miles} {et~al.}(2025{\natexlab{a}}){Miles}, {Shannon}, {Reardon}, {Bailes}, {Champion}, {Geyer}, {Gitika}, {Grunthal}, {Keith}, {Kramer}, {Kulkarni}, {Nathan}, {Parthasarathy}, {Porayko}, {Singha}, {Theureau}, {Abbate}, {Buchner}, {Cameron}, {Camilo}, {Moreschi}, {Shaifullah}, {Shamohammadi}, \& {Krishnan}}]{meerkat_data}
{Miles}, M.~T., {Shannon}, R.~M., {Reardon}, D.~J., {et~al.} 2025{\natexlab{a}}, \mnras, 536, 1467

\bibitem[{{Miles} {et~al.}(2025{\natexlab{b}}){Miles}, {Shannon}, {Reardon}, {Bailes}, {Champion}, {Geyer}, {Gitika}, {Grunthal}, {Keith}, {Kramer}, {Kulkarni}, {Nathan}, {Parthasarathy}, {Singha}, {Theureau}, {Thrane}, {Abbate}, {Buchner}, {Cameron}, {Camilo}, {Moreschi}, {Shaifullah}, {Shamohammadi}, {Possenti}, \& {Krishnan}}]{meerkat_gwb}
{Miles}, M.~T., {Shannon}, R.~M., {Reardon}, D.~J., {et~al.} 2025{\natexlab{b}}, \mnras, 536, 1489

\bibitem[{{NANOGrav Collaboration} {et~al.}(2015){NANOGrav Collaboration}, {Arzoumanian}, {Brazier}, {Burke-Spolaor}, {Chamberlin}, {Chatterjee}, {Christy}, {Cordes}, {Cornish}, {Crowter}, {Demorest}, {Dolch}, {Ellis}, {Ferdman}, {Fonseca}, {Garver-Daniels}, {Gonzalez}, {Jenet}, {Jones}, {Jones}, {Kaspi}, {Koop}, {Lam}, {Lazio}, {Levin}, {Lommen}, {Lorimer}, {Luo}, {Lynch}, {Madison}, {McLaughlin}, {McWilliams}, {Nice}, {Palliyaguru}, {Pennucci}, {Ransom}, {Siemens}, {Stairs}, {Stinebring}, {Stovall}, {Swiggum}, {Vallisneri}, {van Haasteren}, {Wang}, \& {Zhu}}]{nanograv2015}
{NANOGrav Collaboration}, {Arzoumanian}, Z., {Brazier}, A., {et~al.} 2015, \apj, 813, 65

\bibitem[{{Perera} {et~al.}(2019){Perera}, {DeCesar}, {Demorest}, {Kerr}, {Lentati}, {Nice}, {Os{\l}owski}, {Ransom}, {Keith}, {Arzoumanian}, {Bailes}, {Baker}, {Bassa}, {Bhat}, {Brazier}, {Burgay}, {Burke-Spolaor}, {Caballero}, {Champion}, {Chatterjee}, {Chen}, {Cognard}, {Cordes}, {Crowter}, {Dai}, {Desvignes}, {Dolch}, {Ferdman}, {Ferrara}, {Fonseca}, {Goldstein}, {Graikou}, {Guillemot}, {Hazboun}, {Hobbs}, {Hu}, {Islo}, {Janssen}, {Karuppusamy}, {Kramer}, {Lam}, {Lee}, {Liu}, {Luo}, {Lyne}, {Manchester}, {McKee}, {McLaughlin}, {Mingarelli}, {Parthasarathy}, {Pennucci}, {Perrodin}, {Possenti}, {Reardon}, {Russell}, {Sanidas}, {Sesana}, {Shaifullah}, {Shannon}, {Siemens}, {Simon}, {Spiewak}, {Stairs}, {Stappers}, {Swiggum}, {Taylor}, {Theureau}, {Tiburzi}, {Vallisneri}, {Vecchio}, {Wang}, {Zhang}, {Zhang}, {Zhu}, \& {Zhu}}]{ipta_dr2}
{Perera}, B.~B.~P., {DeCesar}, M.~E., {Demorest}, P.~B., {et~al.} 2019, \mnras, 490, 4666

\bibitem[{{Press} {et~al.}(2007){Press}, {Teukolsky}, {Vetterling}, \& {Flannery}}]{NR3}
{Press}, W.~H., {Teukolsky}, S.~A., {Vetterling}, W.~T., \& {Flannery}, B.~P. 2007, {Numerical recipes in C++ : the art of scientific computing} ({Cambridge university press, Cambridge CB2 8RU, UK})

\bibitem[{{Reardon} {et~al.}(2023{\natexlab{a}}){Reardon}, {Zic}, {Shannon}, {Di Marco}, {Hobbs}, {Kapur}, {Lower}, {Mandow}, {Middleton}, {Miles}, {Rogers}, {Askew}, {Bailes}, {Bhat}, {Cameron}, {Kerr}, {Kulkarni}, {Manchester}, {Nathan}, {Russell}, {Os{\l}owski}, \& {Zhu}}]{pptadr3_noise}
{Reardon}, D.~J., {Zic}, A., {Shannon}, R.~M., {et~al.} 2023{\natexlab{a}}, \apjl, 951, L7

\bibitem[{{Reardon} {et~al.}(2023{\natexlab{b}}){Reardon}, {Zic}, {Shannon}, {Hobbs}, {Bailes}, {Di Marco}, {Kapur}, {Rogers}, {Thrane}, {Askew}, {Bhat}, {Cameron}, {Cury{\l}o}, {Coles}, {Dai}, {Goncharov}, {Kerr}, {Kulkarni}, {Levin}, {Lower}, {Manchester}, {Mandow}, {Miles}, {Nathan}, {Os{\l}owski}, {Russell}, {Spiewak}, {Zhang}, \& {Zhu}}]{pptadr3_gwb}
{Reardon}, D.~J., {Zic}, A., {Shannon}, R.~M., {et~al.} 2023{\natexlab{b}}, \apjl, 951, L6

\bibitem[{{Tarafdar} {et~al.}(2022){Tarafdar}, {Nobleson}, {Rana}, {Singha}, {Krishnakumar}, {Joshi}, {Paladi}, {Kolhe}, {Batra}, {Agarwal}, {Bathula}, {Dandapat}, {Desai}, {Dey}, {Hisano}, {Ingale}, {Kato}, {Kharbanda}, {Kikunaga}, {Marmat}, {Pandian}, {Prabu}, {Srivastava}, {Surnis}, {Susarla}, {Susobhanan}, {Takahashi}, {Arumugam}, {Bagchi}, {Banik}, {De}, {Girgaonkar}, {Gopakumar}, {Gupta}, {Maan}, {Manoharan}, {Naidu}, \& {Pathak}}]{inpta_dr1}
{Tarafdar}, P., {Nobleson}, K., {Rana}, P., {et~al.} 2022, \pasa, 39, e053

\bibitem[{{Taylor} {et~al.}(2017){Taylor}, {Lentati}, {Babak}, {Brem}, {Gair}, {Sesana}, \& {Vecchio}}]{tlb+2017}
{Taylor}, S.~R., {Lentati}, L., {Babak}, S., {et~al.} 2017, \prd, 95, 042002

\bibitem[{{Vallisneri}(2020)}]{libstempo}
{Vallisneri}, M. 2020, {libstempo: Python wrapper for Tempo2}, Astrophysics Source Code Library, record ascl:2002.017

\bibitem[{van Haasteren \& Levin(2013)}]{vhl2013}
van Haasteren, R. \& Levin, Y. 2013, \mnras, 428, 1147

\bibitem[{van Haasteren {et~al.}(2009)van Haasteren, Levin, McDonald, \& Lu}]{vhl+2009}
van Haasteren, R., Levin, Y., McDonald, P., \& Lu, T. 2009, \mnras, 395, 1005

\bibitem[{van Haasteren \& Vallisneri(2014)}]{vhv2014}
van Haasteren, R. \& Vallisneri, M. 2014, Phys. Rev. D, 90, 104012

\bibitem[{van Haasteren \& Vallisneri(2015)}]{vhv2015}
van Haasteren, R. \& Vallisneri, M. 2015, \mnras, 446, 1170

\bibitem[{{Verbiest} {et~al.}(2016){Verbiest}, {Lentati}, {Hobbs}, {van Haasteren}, {Demorest}, {Janssen}, {Wang}, {Desvignes}, {Caballero}, {Keith}, {Champion}, {Arzoumanian}, {Babak}, {Bassa}, {Bhat}, {Brazier}, {Brem}, {Burgay}, {Burke-Spolaor}, {Chamberlin}, {Chatterjee}, {Christy}, {Cognard}, {Cordes}, {Dai}, {Dolch}, {Ellis}, {Ferdman}, {Fonseca}, {Gair}, {Garver-Daniels}, {Gentile}, {Gonzalez}, {Graikou}, {Guillemot}, {Hessels}, {Jones}, {Karuppusamy}, {Kerr}, {Kramer}, {Lam}, {Lasky}, {Lassus}, {Lazarus}, {Lazio}, {Lee}, {Levin}, {Liu}, {Lynch}, {Lyne}, {Mckee}, {McLaughlin}, {McWilliams}, {Madison}, {Manchester}, {Mingarelli}, {Nice}, {Os{\l}owski}, {Palliyaguru}, {Pennucci}, {Perera}, {Perrodin}, {Possenti}, {Petiteau}, {Ransom}, {Reardon}, {Rosado}, {Sanidas}, {Sesana}, {Shaifullah}, {Shannon}, {Siemens}, {Simon}, {Smits}, {Spiewak}, {Stairs}, {Stappers}, {Stinebring}, {Stovall}, {Swiggum}, {Taylor}, {Theureau}, {Tiburzi}, {Toomey}, {Vallisneri}, {van Straten}, {Vecchio}, {Wang}, {Wen}, {You},
  {Zhu}, \& {Zhu}}]{ipta_dr1}
{Verbiest}, J.~P.~W., {Lentati}, L., {Hobbs}, G., {et~al.} 2016, \mnras, 458, 1267

\bibitem[{Xu {et~al.}(2023)Xu, Chen, Guo, Jiang, Wang, Xu, Xue, Caballero, Yuan, Xu, Wang, Hao, Luo, Lee, Han, Jiang, Shen, Wang, Wang, Xu, Wu, Manchester, Qian, Guan, Huang, Sun, \& Zhu}]{cpta_dr1}
Xu, H., Chen, S., Guo, Y., {et~al.} 2023, Research in Astronomy and Astrophysics, 23, 075024

\end{thebibliography}

\appendix

\section{Comparison between jitter noise models}
\label{sec:jitter}

We compare two different models for the jitter noise, where, in model I, the white noise is modeled using EQUAD and ECORR as explained in the main text Section \ref{sec:methods}; and, in model II, the white noise is modeled as
\begin{equation}
\begin{split}
{\rm C}_{{\rm WN}, jk} = &\ \EFAC^2 \sigma_j \sigma_k \delta_{jk} + \EQD^2 \left(\frac{\rm 1 hr}{t_{\rm obs}}\right)^2 \delta_{jk}
\\
&+ \ECR^2 \left(\frac{\rm 1 hr}{t_{\rm obs}}\right)^2 {\rm C}_{\ECR,jk} \,,
\end{split}
\end{equation}
i.e. both the uncorrelated jitter (EQUAD) and correlated jitter (ECORR) are modeled with their values being proportional to the observation time length $\sqrt{t_{\rm obs}}$. We list the Bayes factors between the two models in \TAB{tab:jitvsnojit}. As one can see we can not tell the preference between the two models for most of the pulsars as the observation time length of each epoch is roughly a constant. For J1713$+$0747, J2033$+$1734 the time dependent jitter noise model is preferred with $\ln {\cal B}\ge 3$, while the standard EQUAD-ECORR model is only preferred for J1643$-$1224. It is interesting to note that the jitter noise model is not systematically better than the EQUAD-ECORR model.

\begin{table}[h]
\renewcommand{\arraystretch}{1.3}
\caption{Bayesian evidence differences.}
\centering
\begin{tabular}{c|c}
\hline \hline
Pulsar name & $\Delta \ln {\cal B}$ (Model II - I)  \\
\hline
J0023$+$0923 &-1 \\
J0218$+$4232 &1 \\
J0340$+$4130 &1 \\
J0406$+$3039 &1 \\
J0645$+$5158 &-1 \\
J1012$+$5307 &1 \\
J1640$+$2224 &-1 \\
J1643$-$1224 &-6 \\
J1713$+$0747 &8 \\
J1744$-$1134 &-3 \\
J1832$-$0836 &1 \\
J1857$+$0943 &-1 \\
J1910$+$1256 &2 \\
J2010$-$1323 &1 \\
J2017$+$0603 &-1 \\
J2033$+$1734 &8 \\
J2043$+$1711 &-2 \\
J2145$-$0750 &-2 \\
J2214$+$3000 &-1 \\
J2229$+$2643 &-2 \\
J2234$+$0611 &-1 \\
J2234$+$0944 &1 \\
J2302$+$4442 &1 \\
\hline \hline
\end{tabular}
\label{tab:jitvsnojit}
\tablefoot{Bayesian evidence differences between the EQUAD-ECORR model (Model I) and time dependent jitter model (Model II), the higher value (more positive), the larger is the preference for Model II. In this table we omit pulsars with $\Delta |\ln {\cal B}|<1$.}
\end{table}

\section{Numeric integration weights and covariance matrix computation}
\label{sec:romberg}

The covariance matrix ${\rm C}_{j,k}$ with $j$ and $k$ being the indices for time epochs 
and power spectral density function $S(f)$ with $f$ being in the frequency domain 
are related to each other via the Fourier transform
\begin{equation}
        {\rm C}_{ij}=\int S(f) \cos\left (2\pi f(t_i-t_j)\right)\,{\rm d}f\,.
        \label{eq:covspec}
\end{equation}
The above equation can also be computed by integrating over ${\rm d}\ln f$ to get
\begin{equation}
        {\rm C}_{ij}=\int S(f) f \cos\left (2\pi f(t_i-t_j)\right)\,{\rm d}\ln f\,.
        \label{eq:covspeclog}
\end{equation}
In order to speed up the numerical computation the PTA community usually use 
the frequency domain approximation of the covariance matrix, i.e.
\begin{equation}
        {\C}={\bf F S F}^{\rm T}\,,
        \label{eq:freqcov}
\end{equation}
which is a consequence of approximating Eqn.~\ref{eq:covspeclog} using the following summation
\begin{equation}
        {\rm C}_{jk}=\sum_{l} S_l  (c_j c_k + s_j s_k)\,,
        \label{eq:lineexp}
\end{equation}
and casting the equation into the matrix form of Eq~\ref{eq:freqcov}, where 
$c_j\equiv\cos 2\pi f_l t_j$, $s_j\equiv\sin 2\pi f_l t_j$, $f_l$ is the discrete 
frequency, and $S_l= S(f_l) \Delta f$. 
Here, $l$ increases up to $2,3,5,9,...,2^n+1$, where $n$ is the order of integration.
For the integration over ${\rm d}\ln f$, we have $S_l=S(f_l) f_l \Delta \ln f$. However, the approximation 
in Eqn.~\ref{eq:lineexp} is far from being accurate for most applications 
\citep{NR3}. The difference between Eqn.~\ref{eq:covspeclog} and \ref{eq:lineexp} is 
on the order of $O(\Delta f^2)$. One widely used numerical method is to insert 
weights in the summation, such that the difference between the summation and 
integration can be of higher order of $\Delta f$. For sufficiently smooth functions 
the Romberg integration method \citep{NR3} can push the precision to the order of 
$\Delta f^{2n}$ using a slightly modified version of the summation from Eqn.~\ref{eq:lineexp}:
\begin{equation}
        {\rm C}_{jk}=\sum_{l} S_l W_l (c_j c_k + s_j s_k)\,,
        \label{eq:romexp}
\end{equation}
where $W_l$ are the Romberg weights \citep{NR3,NME}. They can be pre-computed using $R(n,m)$ representing Romberg $n$-order integration with $2^n+1$ bins and $m$-order extrapolation ($m\leq n$). For the $l$-th bin, $R_l(n,m)$ is defined recursively as
\begin{equation}
R_{l}(n,0)=
\begin{cases}
\frac{1}{2}, & {\rm for}\, l=1,2^n+1\,; \\
1, & {\rm for}\, l=2,...,2^n \,({\rm if}\, n\geq1)\,;
\end{cases}
\end{equation}
\begin{equation}
R_{l}(n,m)=\frac{1}{4^m-1}[4^m R_{l}(n,m-1)-R_{l}(n-1,m-1)]. 
\end{equation}
Extrapolating $m$ to the $n$-order gives the Romberg weights for the $l$-th bin (up to $2^n+1$), $W_l=R_l(n,n)$.
In 42++, all the covariance matrices are computed with Eqn.~\ref{eq:romexp}. Our numerical 
experiments show that using Romberg weights can improve the precision of ${\rm C}_{jk}$ 
by two to three orders of magnitude for most of the number of frequency bins and spectral 
indices encountered in the application of PTAs, although the improvement of the likelihood evaluation precision can be limited by other factors. More details will be published elsewhere.

\section{Frequency binning and properties of the data set}

\onecolumn

\begin{table}
\renewcommand{\arraystretch}{1.0}
\caption{Frequency binning and properties of the data set.}
\centering
\begin{tabular}{c|cc|cc|cccc}
\hline\hline
DM model& \multicolumn{2}{c|}{DM GP} & \multicolumn{2}{c|}{DMX} & \multicolumn{4}{c}{properties} \\
\hline
Pulsar  &   $n_{\rm bin}$ & $n_{\rm high}$  &   $n_{\rm bin}$ & $n_{\rm high}$  & $T$ [yr] & $n_{\rm epoch}$ & RMS [ns] & $\Delta {\rm DMX} \, [\times 10^{-5} {\rm pc} \, {\rm cm}^{-3}]$\\
\hline
J0023$+$0923  &      $65$                 &$65$                        &        $33$               &$35$               &2.6&  61 & 191  & 18.1\\
J0030$+$0451  &      $65$                 &$58$                        &        $33$               &$18$               &2.7&  61 & 96   & 4.6 \\
J0034$-$0534  &      $33$                 &$56$                        &        $33$               &$17$               &2.6&  49 & 240  & 116.0\\
J0154$+$1833  &      $33$                 &$18$                        &        $33$               &$14$               &2.2&  29 & 32   & 18.3\\
J0218$+$4232  &      $65$                 &$74$                        &        $65$               &$84$               &3.4& 123 & 182  & 6.8 \\
J0340$+$4130  &      $65$                 &$83$                        &        $33$               &$16$               &3.2&  70 & 95   & 24.0\\
J0406$+$3039  &      $33$                 &$25$                        &        $33$               &$16$               &1.2&  30 & 38   & 35.4\\
J0509$+$0856  &      $33$                 &$16$                        &        $33$               &$18$               &2.4&  37 & 171  & 18.8\\
J0605$+$3757  &      $33$                 &$14$                        &        $33$               &$17$               &2.4&  24 & 403  & 49.6\\
J0613$-$0200  &      $65$                 &$86$                        &        $33$               &$15$               &3.2&  67 & 47   & 11.3\\
J0621$+$1002  &      $129$                &$83$                        &        $33$               &$18$               &3.2&  69 & 488  & 25.3\\
J0636$+$5128  &      $65$                 &$178$                       &        $65$               &$114$              &3.2&  71 & 203  & 8.9 \\
J0645$+$5158  &      $65$                 &$23$                        &        $33$               &$17$               &3.2&  69 & 39   & 12.8\\
J0732$+$2334  &      $33$                 &$18$                        &        $33$               &$16$               &2.4&  23 & 244  & 33.9\\
J0751$+$1807  &      $65$                 &$63$                        &        $33$               &$15$               &3.3&  70 & 107  & 3.2 \\
J0824$+$0028  &      $33$                 &$17$                        &        $33$               &$18$               &2.4&  23 & 277  & 45.0\\
J1012$+$5307  &      $65$                 &$55$                        &        $33$               &$32$               &3.2&  69 & 65   & 5.5 \\
J1024$-$0719  &      $33$                 &$59$                        &        $33$               &$19$               &2.4&  65 & 51   & 8.2 \\
J1327$+$3423  &      $33$                 &$11$                        &        $33$               &$11$               &1.1&  28 & 114  & 48.2\\
J1453$+$1902  &      $33$                 &$16$                        &        $33$               &$16$               &3.2&  63 & 258  & 41.7\\
J1630$+$3734  &      $33$                 &$18$                        &        $33$               &$16$               &2.6&  47 & 44   & 11.7\\
J1640$+$2224  &      $129$                &$178$                       &        $33$               &$61$               &3.1&  69 & 44   & 5.9 \\
J1643$-$1224  &      $65$                 &$144$                       &        $65$               &$68$               &2.6&  56 & 295  & 7.5 \\
J1710$+$4923  &      $33$                 &$19$                        &        $33$               &$15$               &2.4&  21 & 196  & 17.4\\
J1713$+$0747  &      $65$                 &$112$                       &                           &                   &1.7&  66 & 63   & 2.6 \\
J1738$+$0333  &      $129$                &$73$                        &        $65$               &$66$               &3.2&  65 & 108  & 5.3 \\
J1741$+$1351  &      $129$                &$102$                       &        $33$               &$15$               &3.2&  67 & 57   & 4.2 \\
J1744$-$1134  &      $33$                 &$84$                        &        $33$               &$14$               &2.6&  59 & 268  & 2.1 \\
J1745$+$1017  &      $65$                 &$47$                        &        $33$               &$17$               &2.4&  20 & 497  & 16.2\\
J1832$-$0836  &      $33$                 &$68$                        &        $33$               &$19$               &2.6&  52 & 46   & 4.6 \\
J1843$-$1113  &      $65$                 &$73$                        &        $33$               &$16$               &2.6&  47 & 38   & 15.3\\
J1853$+$1303  &      $33$                 &$15$                        &        $33$               &$17$               &3.1&  53 & 71   & 8.6 \\
J1857$+$0943  &      $33$                 &$59$                        &        $33$               &$15$               &3.1&  55 & 129  & 4.5 \\
J1903$+$0327  &      $33$                 &$74$                        &        $33$               &$23$               &2.6&  50 & 115  & 92.7\\
J1910$+$1256  &      $33$                 &$76$                        &        $33$               &$15$               &3.1&  51 & 46   & 5.9 \\
J1911$-$1114  &      $33$                 &$29$                        &        $33$               &$15$               &2.6&  58 & 195  & 14.9\\
J1911$+$1347  &      $33$                 &$15$                        &        $33$               &$14$               &3.1&  53 & 82   & 2.3 \\
J1918$-$0642  &      $65$                 &$27$                        &        $33$               &$15$               &2.6&  51 & 60   & 4.5 \\
J1923$+$2515  &      $33$                 &$15$                        &        $33$               &$17$               &3.1&  55 & 141  & 12.9\\
J1944$+$0907  &      $33$                 &$14$                        &        $33$               &$18$               &1.8&  35 & 362  & 33.9\\
J1946$+$3417  &      $65$                 &$23$                        &        $65$               &$112$              &2.4&  58 & 155  & 12.0\\
J1955$+$2908  &      $65$                 &$62$                        &        $33$               &$19$               &2.6&  55 & 191  & 17.6\\
J2010$-$1323  &      $65$                 &$57$                        &        $33$               &$16$               &2.6&  47 & 51   & 5.8 \\
J2017$+$0603  &      $33$                 &$15$                        &        $33$               &$15$               &2.6&  54 & 104  & 9.8 \\
J2019$+$2425  &      $33$                 &$16$                        &        $33$               &$17$               &2.6&  57 & 284  & 22.3\\
J2022$+$2534  &      $33$                 &$16$                        &        $33$               &$14$               &1.2&  27 & 51   & 9.0 \\
J2033$+$1734  &      $65$                 &$117$                       &        $33$               &$15$               &2.6&  57 & 129  & 34.3\\
J2043$+$1711  &      $33$                 &$13$                        &        $33$               &$104$              &2.6&  65 & 121  & 15.1\\
J2145$-$0750  &      $33$                 &$17$                        &        $33$               &$16$               &2.6&  64 & 277  & 14.8\\
J2150$-$0326  &      $33$                 &$7$                         &        $33$               &$15$               &1.2&  33 & 183  & 10.7\\
J2214$+$3000  &      $33$                 &$39$                        &        $33$               &$77$               &2.7&  62 & 181  & 10.1\\
J2229$+$2643  &      $33$                 &$71$                        &        $33$               &$62$               &2.6&  65 & 84   & 14.2\\
J2234$+$0611  &      $33$                 &$14$                        &        $33$               &$14$               &2.6&  68 & 99   & 6.7 \\
J2234$+$0944  &      $65$                 &$83$                        &        $33$               &$72$               &2.6&  66 & 55   & 6.7 \\
J2302$+$4442  &      $33$                 &$73$                        &        $33$               &$13$               &2.5&  64 & 177  & 10.0\\
J2337$+$1439  &      $65$                 &$59$                        &        $33$               &$16$               &2.6&  70 & 83   & 7.1 \\
J2322$+$2057  &      $33$                 &$44$                        &        $33$               &$16$               &2.6&  65 & 85   & 12.7\\
\hline\hline
\end{tabular}
\label{tab:freq}
\tablefoot{Optimal choice of number of frequency bins $n_{\rm bin}$ and high frequency cutoff number $n_{\rm high}$ for the power law model for the red noise and DM GP in the Bayesian analysis for the DM GP and DMX data sets; as well as the time span $T$, number of epoch $n_{\rm epoch}$, (whitened and band averaged) residual RMS and DMX value uncertainties $\Delta {\rm DMX}$.}
\end{table}

\clearpage
\section{Posterior constraints on the noise parameters}
\label{sec:tab}

\setlength{\tabcolsep}{5.0pt}
\renewcommand{\arraystretch}{1.3}
\begin{center}
{\scriptsize
\begin{longtable}{c|ccc|cc|cc|ccc}
\caption{Summary of the noise parameters of the CPTA MSPs recovered using \texttt{TEMPONEST}.}
\label{tab:tn}
\\\hline \hline
DM model & \multicolumn{3}{c|}{DM GP} & \multicolumn{2}{c|}{DM GP} & \multicolumn{2}{c|}{DMX} & \multicolumn{3}{c}{DM GP} \\
\hline
Pulsar name & EFAC & $\log_{10}$ EQUAD & $\log_{10}$ ECORR & $\log_{10}(A_{\rm RN})$ & $\gamma_{\rm RN}$ & $\log_{10}(A_{\rm RN})$ & $\gamma_{\rm RN}$ & $\log_{10}(A_{\rm DM})$ & $\gamma_{\rm DM}$ & $\log_{10}(A_{\rm yr})$ \\
\hline 
J0023$+$0923  &$1.047 \pm 0.006$            &$< -7.41$                    &$-6.95 \pm 0.06$             &$-13.44 \,^{+0.21}_{-0.15}$  &$0.43 \,^{+0.40}_{-0.31}$    &$-13.38 \,^{+0.29}_{-0.40}$  &$0.7 \,^{+1.5}_{-0.5}$      &$< -11.51$                   &--                           &$< -3.92$                    \\
J0030$+$0451  &$0.968 \pm 0.010$            &$< -7.28$                    &$-6.96 \,^{+0.13}_{-0.11}$   &$< -13.17$                   &--                           &$< -13.01$                   &--                          &$< -11.78$                   &--                           &$< -4.22$                    \\
J0034$-$0534  &$1.035 \,^{+0.016}_{-0.015}$ &$< -6.78$                    &$< -6.42$                    &$< -12.36$                   &--                           &$< -11.91$                   &--                          &$< -11.06$                   &--                           &$< -3.29$                    \\
J0154$+$1833  &$0.989 \pm 0.018$            &$< -7.11$                    &$< -6.27$                    &$< -12.21$                   &--                           &$< -12.26$                   &--                          &$< -11.06$                   &--                           &$< -3.60$                    \\
J0218$+$4232  &$0.650 \pm 0.007$            &$< -6.92$                    &$-6.91 \pm 0.05$             &$-13.04 \,^{+0.12}_{-0.13}$  &$1.66 \,^{+0.40}_{-0.32}$    &$-13.01 \,^{+0.15}_{-0.17}$  &$0.74 \,^{+0.34}_{-0.40}$   &$-11.26 \pm 0.08$            &$2.09 \,^{+0.25}_{-0.21}$    &$< -3.53$                    \\
J0340$+$4130  &$1.019 \,^{+0.016}_{-0.017}$ &$-6.50 \,^{+0.04}_{-0.05}$   &$< -7.10$                    &$< -13.34$                   &--                           &$< -12.56$                   &--                          &$-11.51 \,^{+0.11}_{-0.12}$  &$1.94 \,^{+0.32}_{-0.30}$    &$-3.31 \,^{+0.05}_{-0.07}$   \\
J0406$+$3039  &$0.91 \pm 0.04$              &$-6.346 \,^{+0.031}_{-0.040}$&$< -6.70$                    &$< -11.61$                   &--                           &$< -11.07$                   &--                          &$-11.28 \,^{+0.25}_{-0.26}$  &$1.7 \pm 0.6$                &$< -3.22$                    \\
J0509$+$0856  &$1.061 \,^{+0.018}_{-0.022}$ &$< -6.35$                    &$< -6.62$                    &$< -12.81$                   &--                           &$< -12.19$                   &--                          &$< -11.40$                   &--                           &$< -3.70$                    \\
J0605$+$3757  &$1.060 \,^{+0.017}_{-0.016}$ &$< -6.57$                    &$< -6.25$                    &$< -12.14$                   &--                           &$< -11.88$                   &--                          &$< -11.05$                   &--                           &$< -3.32$                    \\
J0613$-$0200  &$0.957 \pm 0.010$            &$< -7.18$                    &$-7.24 \,^{+0.13}_{-0.16}$   &$< -13.18$                   &--                           &$< -13.07$                   &--                          &$-11.70 \,^{+0.15}_{-0.17}$  &$3.0 \,^{+1.3}_{-0.8}$       &$< -3.66$                    \\
J0621$+$1002  &$0.939 \pm 0.010$            &$< -6.89$                    &$-5.86 \,^{+0.09}_{-0.18}$   &$< -12.19$                   &--                           &$< -12.24$                   &--                          &$-10.74 \,^{+0.09}_{-0.08}$  &$2.00 \,^{+0.24}_{-0.22}$    &$< -2.97$                    \\
J0636$+$5128  &$1.051 \pm 0.007$            &$-7.52 \,^{+0.05}_{-0.06}$   &$-6.831 \,^{+0.018}_{-0.019}$&$< -13.54$                   &--                           &$< -13.30$                   &--                          &$-12.49 \,^{+0.17}_{-0.14}$  &$0.41 \,^{+0.40}_{-0.28}$    &$< -4.43$                    \\
J0645$+$5158  &$1.048 \,^{+0.011}_{-0.010}$ &$-7.58 \,^{+0.10}_{-0.13}$   &$< -6.84$                    &$< -13.33$                   &--                           &$< -12.64$                   &--                          &$< -12.05$                   &--                           &$< -4.28$                    \\
J0732$+$2314  &$1.083 \,^{+0.024}_{-0.021}$ &$< -6.58$                    &$< -6.29$                    &$< -12.27$                   &--                           &$< -11.66$                   &--                          &$< -11.35$                   &--                           &$< -3.69$                    \\
J0751$+$1807  &$1.001 \pm 0.006$            &$< -7.38$                    &$-7.03 \pm 0.04$             &$< -13.43$                   &--                           &$-13.91 \,^{+0.35}_{-1.00}$  &$2.5 \,^{+3.1}_{-1.9}$      &$-11.85 \,^{+0.12}_{-0.13}$  &$1.35 \pm 0.31$              &$< -4.04$                    \\
J0824$+$0028  &$1.030 \,^{+0.021}_{-0.023}$ &$< -6.14$                    &$< -6.11$                    &$< -12.34$                   &--                           &$< -12.05$                   &--                          &$< -11.04$                   &--                           &$< -3.27$                    \\
J1012$+$5307  &$0.921 \pm 0.009$            &$< -7.45$                    &$-7.09 \pm 0.10$             &$-13.06 \,^{+0.16}_{-0.15}$  &$1.58 \,^{+0.40}_{-0.34}$    &$-12.97 \,^{+0.18}_{-0.17}$  &$1.4 \pm 0.4$               &$< -11.89$                   &--                           &$< -4.14$                    \\
J1024$-$0719  &$1.024 \pm 0.011$            &$< -7.04$                    &$< -7.04$                    &$< -13.00$                   &--                           &$< -12.07$                   &--                          &$-12.37 \,^{+0.21}_{-0.23}$  &$0.8 \pm 0.5$                &$< -4.27$                    \\
J1327$+$3423  &$1.055 \,^{+0.019}_{-0.018}$ &$< -6.77$                    &$< -5.64$                    &$< -11.01$                   &--                           &$-11.8 \,^{+0.6}_{-0.8}$     &$2.5 \,^{+1.5}_{-1.2}$      &$< -10.93$                   &--                           &$< -3.62$                    \\
J1453$+$1902  &$1.059 \pm 0.011$            &$< -6.58$                    &$< -6.51$                    &$< -12.72$                   &--                           &$< -12.56$                   &--                          &$< -11.43$                   &--                           &$< -3.80$                    \\
J1630$+$3734  &$1.013 \pm 0.015$            &$< -7.04$                    &$< -6.76$                    &$< -12.85$                   &--                           &$< -12.65$                   &--                          &$< -11.48$                   &--                           &$< -4.11$                    \\
J1640$+$2224  &$1.018 \pm 0.010$            &$-7.27 \pm 0.05$             &$-7.36 \,^{+0.10}_{-0.12}$   &$-13.24 \,^{+0.20}_{-0.16}$  &$0.47 \,^{+0.40}_{-0.30}$    &$-12.98 \,^{+0.20}_{-0.24}$  &$1.1 \pm 0.5$               &$-12.11 \,^{+0.17}_{-0.15}$  &$0.31 \,^{+0.35}_{-0.23}$    &$< -4.19$                    \\
J1643$-$1224  &$1.14 \pm 0.07$              &$-6.328 \,^{+0.020}_{-0.023}$&$< -6.10$                    &$-12.37 \,^{+0.19}_{-0.16}$  &$1.9 \,^{+0.8}_{-0.5}$       &$-12.32 \,^{+0.16}_{-0.15}$  &$2.0 \pm 0.4$               &$-10.83 \pm 0.16$            &$2.2 \pm 0.5$                &$< -2.82$                    \\
J1710$+$4923  &$1.006 \,^{+0.021}_{-0.020}$ &$< -6.62$                    &$< -6.44$                    &$< -12.94$                   &--                           &$< -12.49$                   &--                          &$< -11.39$                   &--                           &$< -3.65$                    \\
J1713$+$0747  &$0.831 \pm 0.006$            &$-7.852 \pm 0.022$           &$-7.163 \,^{+0.024}_{-0.023}$&$-13.6 \pm 0.4$              &$1.8 \,^{+0.9}_{-0.8}$       &$-13.6 \,^{+0.7}_{-1.1}$     &$2.5 \,^{+2.0}_{-1.4}$      &$< -11.97$                   &--                           &$< -4.54$                    \\
J1738$+$0333  &$1.030 \pm 0.008$            &$< -7.31$                    &$-7.02 \pm 0.06$             &$< -12.96$                   &--                           &$-13.45 \,^{+0.22}_{-0.18}$  &$0.47 \,^{+0.50}_{-0.33}$   &$-11.14 \,^{+0.10}_{-0.09}$  &$2.82 \,^{+0.35}_{-0.32}$    &$< -3.29$                    \\
J1741$+$1351  &$1.015 \,^{+0.011}_{-0.010}$ &$< -7.43$                    &$-7.21 \,^{+0.10}_{-0.09}$   &$< -13.38$                   &--                           &$< -13.36$                   &--                          &$-12.05 \,^{+0.12}_{-0.13}$  &$0.97 \,^{+0.31}_{-0.28}$    &$< -4.23$                    \\
J1744$-$1134  &$1.094 \pm 0.019$            &$-7.094 \,^{+0.019}_{-0.021}$&$-6.53 \pm 0.05$             &$< -13.15$                   &--                           &$< -13.21$                   &--                          &$-12.38 \,^{+0.18}_{-0.14}$  &$0.36 \,^{+0.40}_{-0.25}$    &$< -4.50$                    \\
J1745$+$1017  &$1.081 \,^{+0.025}_{-0.027}$ &$< -6.68$                    &$< -5.68$                    &$< -11.49$                   &--                           &$< -11.53$                   &--                          &$-11.08 \pm 0.23$            &$0.8 \,^{+0.6}_{-0.5}$       &$< -3.00$                    \\
J1832$-$0836  &$0.969 \pm 0.011$            &$< -7.36$                    &$< -7.07$                    &$< -13.08$                   &--                           &$< -13.12$                   &--                          &$-11.24 \,^{+0.09}_{-0.08}$  &$2.82 \,^{+0.34}_{-0.27}$    &$< -3.49$                    \\
J1843$-$1113  &$1.002 \,^{+0.012}_{-0.013}$ &$< -7.08$                    &$< -7.02$                    &$< -13.06$                   &--                           &$< -12.91$                   &--                          &$-11.29 \pm 0.09$            &$2.34 \,^{+0.27}_{-0.25}$    &$< -3.47$                    \\
J1853$+$1303  &$1.022 \,^{+0.011}_{-0.012}$ &$< -7.30$                    &$< -7.12$                    &$< -13.17$                   &--                           &$< -13.02$                   &--                          &$< -11.81$                   &--                           &$< -4.20$                    \\
J1857$+$0943  &$0.966 \pm 0.015$            &$-7.25 \,^{+0.08}_{-0.11}$   &$< -6.59$                    &$< -13.21$                   &--                           &$< -12.98$                   &--                          &$-11.73 \pm 0.11$            &$1.63 \,^{+0.40}_{-0.34}$    &$< -3.88$                    \\
J1903$+$0327  &$1.142 \pm 0.013$            &$< -6.55$                    &$< -6.33$                    &$-11.99 \,^{+0.16}_{-0.14}$  &$4.3 \,^{+1.0}_{-0.9}$       &$-11.90 \,^{+0.17}_{-0.16}$  &$4.1 \,^{+0.9}_{-0.8}$      &$-10.57 \,^{+0.10}_{-0.09}$  &$2.58 \,^{+0.40}_{-0.28}$    &$< -2.84$                    \\
J1910$+$1256  &$1.006 \pm 0.016$            &$-6.89 \,^{+0.06}_{-0.07}$   &$< -6.97$                    &$< -13.22$                   &--                           &$< -13.17$                   &--                          &$-11.71 \pm 0.11$            &$1.91 \,^{+0.40}_{-0.34}$    &$< -3.84$                    \\
J1911$-$1114  &$1.055 \pm 0.013$            &$< -6.64$                    &$< -6.65$                    &$< -12.92$                   &--                           &$< -12.48$                   &--                          &$-11.35 \,^{+0.13}_{-0.12}$  &$3.0 \,^{+0.8}_{-0.6}$       &$< -3.45$                    \\
J1911$+$1347  &$1.021 \,^{+0.014}_{-0.015}$ &$< -7.21$                    &$-7.08 \,^{+0.05}_{-0.06}$   &$< -13.77$                   &--                           &$< -13.44$                   &--                          &$-12.05 \,^{+0.13}_{-0.16}$  &$3.5 \,^{+1.4}_{-1.2}$       &$< -4.11$                    \\
J1918$-$0642  &$1.024 \pm 0.012$            &$< -7.26$                    &$-7.06 \,^{+0.13}_{-0.15}$   &$< -12.93$                   &--                           &$< -12.82$                   &--                          &$-11.96 \,^{+0.17}_{-0.40}$  &$2.3 \,^{+2.9}_{-0.7}$       &$< -3.97$                    \\
J1923$+$2515  &$1.014 \,^{+0.012}_{-0.011}$ &$< -7.09$                    &$< -6.80$                    &$< -13.06$                   &--                           &$< -13.00$                   &--                          &$< -11.56$                   &--                           &$< -3.79$                    \\
J1944$+$0907  &$1.024 \pm 0.017$            &$< -6.99$                    &$< -6.26$                    &$< -12.06$                   &--                           &$< -11.57$                   &--                          &$< -11.11$                   &--                           &$< -3.68$                    \\
J1946$+$3417  &$0.985 \,^{+0.011}_{-0.012}$ &$< -6.62$                    &$-6.63 \,^{+0.07}_{-0.08}$   &$-11.66 \,^{+0.19}_{-0.17}$  &$2.7 \pm 0.5$                &$-11.72 \,^{+0.20}_{-0.17}$  &$2.4 \pm 0.5$               &$-10.08 \,^{+0.05}_{-0.08}$  &$2.81 \,^{+0.20}_{-0.24}$    &$< -2.43$                    \\
J1955$+$2908  &$1.001 \,^{+0.025}_{-0.021}$ &$-6.38 \,^{+0.09}_{-0.22}$   &$< -6.50$                    &$< -12.55$                   &--                           &$< -12.16$                   &--                          &$-11.18 \pm 0.12$            &$2.1 \,^{+0.5}_{-0.4}$       &$< -3.34$                    \\
J2010$-$1323  &$1.068 \pm 0.017$            &$-6.76 \,^{+0.06}_{-0.07}$   &$< -6.73$                    &$-13.34 \,^{+0.19}_{-0.21}$  &$1.2 \pm 0.5$                &$< -12.81$                   &--                          &$< -11.58$                   &--                           &$< -3.99$                    \\
J2017$+$0603  &$1.009 \pm 0.012$            &$< -7.43$                    &$< -6.98$                    &$< -13.16$                   &--                           &$< -12.75$                   &--                          &$-11.80 \,^{+0.16}_{-0.19}$  &$4.5 \,^{+1.4}_{-1.3}$       &$< -3.83$                    \\
J2019$+$2425  &$1.070 \pm 0.012$            &$< -6.67$                    &$-6.72 \,^{+0.10}_{-0.12}$   &$< -13.07$                   &--                           &$< -11.93$                   &--                          &$< -11.68$                   &--                           &$< -3.80$                    \\
J2022$+$2534  &$1.053 \pm 0.019$            &$< -6.92$                    &$< -6.93$                    &$< -11.90$                   &--                           &$< -11.23$                   &--                          &$< -11.13$                   &--                           &$< -3.82$                    \\
J2033$+$1734  &$1.040 \,^{+0.015}_{-0.014}$ &$-6.41 \pm 0.04$             &$< -6.57$                    &$< -13.22$                   &--                           &$< -12.61$                   &--                          &$< -11.82$                   &--                           &$-3.85 \,^{+0.14}_{-0.40}$   \\
J2043$+$1711  &$1.026 \pm 0.014$            &$-6.65 \pm 0.04$             &$< -7.06$                    &$< -12.90$                   &--                           &$-12.94 \,^{+0.23}_{-0.19}$  &$0.5 \,^{+0.5}_{-0.4}$      &$-11.74 \,^{+0.16}_{-0.17}$  &$4.0 \,^{+1.3}_{-1.1}$       &$< -3.69$                    \\
J2145$-$0750  &$0.863 \pm 0.011$            &$-7.26 \,^{+0.06}_{-0.07}$   &$-6.56 \pm 0.05$             &$< -13.00$                   &--                           &$< -12.68$                   &--                          &$< -11.55$                   &--                           &$< -4.02$                    \\
J2150$-$0326  &$1.002 \,^{+0.019}_{-0.018}$ &$-7.14 \,^{+0.11}_{-0.16}$   &$< -6.39$                    &$< -11.64$                   &--                           &$< -11.19$                   &--                          &$-11.23 \,^{+0.34}_{-0.32}$  &$3.2 \,^{+1.6}_{-1.2}$       &$< -2.98$                    \\
J2214$+$3000  &$1.051 \pm 0.013$            &$-6.625 \,^{+0.028}_{-0.029}$&$-6.74 \,^{+0.09}_{-0.10}$   &$< -12.85$                   &--                           &$-13.10 \,^{+0.21}_{-0.17}$  &$0.39 \,^{+0.50}_{-0.28}$   &$-11.53 \,^{+0.12}_{-0.13}$  &$2.1 \,^{+1.0}_{-0.4}$       &$< -3.73$                    \\
J2229$+$2643  &$1.068 \pm 0.013$            &$-6.621 \,^{+0.027}_{-0.030}$&$< -6.70$                    &$-13.63 \,^{+0.28}_{-0.25}$  &$0.7 \,^{+0.7}_{-0.5}$       &$-12.92 \,^{+0.21}_{-0.18}$  &$0.51 \,^{+0.50}_{-0.35}$   &$-12.19 \,^{+0.33}_{-0.32}$  &$5.4 \,^{+1.2}_{-1.7}$       &$< -3.90$                    \\
J2234$+$0611  &$1.060 \pm 0.012$            &$-7.15 \,^{+0.05}_{-0.06}$   &$-7.14 \,^{+0.07}_{-0.08}$   &$< -13.44$                   &--                           &$< -12.83$                   &--                          &$-12.59 \,^{+0.32}_{-0.40}$  &$5.0 \,^{+1.4}_{-1.7}$       &$< -4.13$                    \\
J2234$+$0944  &$0.988 \pm 0.011$            &$-7.18 \,^{+0.05}_{-0.06}$   &$< -6.95$                    &$-12.99 \pm 0.13$            &$1.70 \,^{+0.33}_{-0.30}$    &$-13.19 \,^{+0.21}_{-0.17}$  &$0.47 \,^{+0.50}_{-0.32}$   &$-11.48 \,^{+0.13}_{-0.12}$  &$3.1 \,^{+0.8}_{-0.6}$       &$< -3.60$                    \\
J2302$+$4442  &$1.047 \pm 0.012$            &$< -6.76$                    &$-6.69 \,^{+0.08}_{-0.09}$   &$< -12.86$                   &--                           &$< -12.68$                   &--                          &$-11.52 \pm 0.13$            &$2.2 \,^{+1.0}_{-0.4}$       &$< -3.78$                    \\
J2317$+$1439  &$1.047 \,^{+0.010}_{-0.011}$ &$< -7.21$                    &$-7.08 \pm 0.19$             &$< -13.30$                   &--                           &$< -12.61$                   &--                          &$< -11.59$                   &--                           &$< -3.92$                    \\
J2322$+$2057  &$1.028 \,^{+0.012}_{-0.011}$ &$< -6.91$                    &$< -6.78$                    &$< -13.25$                   &--                           &$< -12.59$                   &--                          &$< -11.54$                   &--                           &$-3.89 \,^{+0.12}_{-0.24}$   \\
\hline\hline
\end{longtable}}
\tablefoot{Detections are given with $1\sigma$ uncertainties, while upper limits are set at 95\%.}
\end{center}

\setlength{\tabcolsep}{5.0pt}
\renewcommand{\arraystretch}{1.3}
\begin{center}
{\scriptsize
\begin{longtable}{c|ccc|cc|cc|ccc}
\caption{Summary of the noise parameters of the CPTA MSPs recovered using \texttt{ENTERPRISE}.}
\label{tab:ep}
\\\hline \hline
DM model & \multicolumn{3}{c|}{DM GP} & \multicolumn{2}{c|}{DM GP} & \multicolumn{2}{c|}{DMX} & \multicolumn{3}{c}{DM GP} \\
\hline
Pulsar name & EFAC & $\log_{10}$ EQUAD & $\log_{10}$ ECORR & $\log_{10}(A_{\rm RN})$ & $\gamma_{\rm RN}$ & $\log_{10}(A_{\rm RN})$ & $\gamma_{\rm RN}$ & $\log_{10}(A_{\rm DM})$ & $\gamma_{\rm DM}$ & $\log_{10}(A_{\rm yr})$ \\
\hline 
J0023$+$0923  &$1.047 \pm 0.006$           &$< -7.37$                   &$-6.93 \pm 0.06$            & $-13.28 \pm 0.17$           &$0.8 \pm 0.4$               &$-13.25 \,^{+0.28}_{-1.50}$ &$1.0 \,^{+2.5}_{-0.7}$      & $< -11.53$                  &--                          &$< -3.94$                   \\
J0030$+$0451  &$0.968 \pm 0.010$           &$< -7.22$                   &$-6.94 \pm 0.10$            & $< -13.24$                  &--                          &$< -13.11$                  &--                          & $< -11.84$                  &--                          &$< -4.22$                   \\
J0034$-$0534  &$1.034 \pm 0.015$           &$< -6.77$                   &$< -6.46$                   & $< -12.33$                  &--                          &$< -11.99$                  &--                          & $< -11.03$                  &--                          &$< -3.32$                   \\
J0154$+$1833  &$0.988 \,^{+0.020}_{-0.019}$&$< -7.10$                   &$-6.46 \,^{+0.11}_{-0.18}$  & $< -12.02$                  &--                          &$< -12.13$                  &--                          & $< -11.05$                  &--                          &$< -3.61$                   \\
J0218$+$4232  &$0.650 \,^{+0.006}_{-0.007}$&$< -6.90$                   &$-6.86 \pm 0.04$            & $-13.11 \,^{+0.14}_{-0.15}$ &$1.3 \pm 0.4$               &$-12.91 \pm 0.14$           &$1.09 \,^{+0.40}_{-0.34}$   & $-11.28 \,^{+0.09}_{-0.08}$ &$1.99 \,^{+0.30}_{-0.25}$   &$< -3.53$                   \\
J0340$+$4130  &$1.019 \,^{+0.017}_{-0.016}$&$-6.50 \pm 0.05$            &$< -7.07$                   & $< -13.30$                  &--                          &$-13.11 \,^{+0.20}_{-1.00}$ &$0.7 \,^{+1.2}_{-0.5}$      & $-11.50 \,^{+0.12}_{-0.11}$ &$1.91 \,^{+0.34}_{-0.30}$   &$-3.31 \,^{+0.06}_{-0.07}$  \\
J0406$+$3039  &$0.91 \pm 0.04$             &$-6.347 \,^{+0.032}_{-0.040}$&$< -6.67$                  & $< -11.37$                  &--                          &$< -11.15$                  &--                          & $-11.21 \,^{+0.23}_{-0.22}$ &$1.9 \pm 0.6$               &$< -3.06$                   \\
J0509$+$0856  &$1.058 \,^{+0.019}_{-0.024}$&$< -6.33$                   &$< -6.61$                   & $< -12.78$                  &--                          &$< -12.13$                  &--                          & $< -11.59$                  &--                          &$< -3.77$                   \\
J0605$+$3757  &$1.059 \,^{+0.017}_{-0.016}$&$< -6.56$                   &$< -6.26$                   & $< -12.08$                  &--                          &$< -11.83$                  &--                          & $< -11.08$                  &--                          &$< -3.32$                   \\
J0613$-$0200  &$0.957 \pm 0.010$           &$< -7.17$                   &$-7.04 \,^{+0.07}_{-0.10}$  & $< -13.16$                  &--                          &$< -13.07$                  &--                          & $-11.78 \,^{+0.17}_{-0.20}$ &$3.6 \,^{+1.5}_{-1.0}$      &$-3.83 \,^{+0.10}_{-0.18}$  \\
J0621$+$1002  &$0.939 \pm 0.010$           &$< -6.87$                   &$-5.90 \,^{+0.12}_{-0.15}$  & $< -12.21$                  &--                          &$< -12.24$                  &--                          & $-10.75 \,^{+0.09}_{-0.08}$ &$1.98 \,^{+0.25}_{-0.23}$   &$< -2.99$                   \\
J0636$+$5128  &$1.051 \pm 0.007$           &$-7.52 \pm 0.05$            &$-6.826 \pm 0.019$          & $< -13.45$                  &--                          &$< -13.43$                  &--                          & $-12.50 \,^{+0.19}_{-0.21}$ &$0.59 \,^{+0.40}_{-0.34}$   &$< -4.38$                   \\
J0645$+$5158  &$1.047 \pm 0.011$           &$-7.57 \,^{+0.10}_{-0.13}$  &$-6.96 \,^{+0.08}_{-0.10}$  & $< -13.24$                  &--                          &$< -12.73$                  &--                          & $< -12.15$                  &--                          &$< -4.29$                   \\
J0732$+$2314  &$1.083 \pm 0.022$           &$< -6.57$                   &$< -6.27$                   & $< -12.27$                  &--                          &$< -11.63$                  &--                          & $< -11.24$                  &--                          &$< -3.66$                   \\
J0751$+$1807  &$1.001 \pm 0.006$           &$< -7.39$                   &$-7.01 \pm 0.04$            & $< -13.42$                  &--                          &$< -13.29$                  &--                          & $-11.82 \,^{+0.12}_{-0.11}$ &$1.33 \,^{+0.32}_{-0.31}$   &$< -4.02$                   \\
J0824$+$0028  &$1.029 \,^{+0.023}_{-0.024}$&$< -6.12$                   &$< -6.08$                   & $< -12.34$                  &--                          &$< -11.94$                  &--                          & $< -11.04$                  &--                          &$< -3.25$                   \\
J1012$+$5307  &$0.921 \pm 0.009$           &$< -7.46$                   &$-7.06 \,^{+0.09}_{-0.10}$  & $-13.14 \,^{+0.19}_{-0.18}$ &$1.3 \pm 0.5$               &$-13.08 \,^{+0.19}_{-0.18}$ &$1.2 \pm 0.6$               & $< -11.92$                  &--                          &$< -4.14$                   \\
J1024$-$0719  &$1.023 \,^{+0.012}_{-0.011}$&$< -7.03$                   &$< -7.05$                   & $< -12.97$                  &--                          &$< -12.13$                  &--                          & $-12.24 \,^{+0.18}_{-0.20}$ &$1.1 \pm 0.5$               &$< -4.25$                   \\
J1327$+$3423  &$1.055 \pm 0.019$           &$< -6.78$                   &$< -6.12$                   & $-12.48 \,^{+0.31}_{-0.19}$ &$0.6 \,^{+0.7}_{-0.4}$      &$-12.1 \,^{+0.5}_{-0.4}$    &$1.3 \,^{+1.4}_{-0.9}$      & $< -11.00$                  &--                          &$< -3.59$                   \\
J1453$+$1902  &$1.059 \pm 0.011$           &$< -6.58$                   &$< -6.53$                   & $< -12.78$                  &--                          &$< -12.65$                  &--                          & $< -11.49$                  &--                          &$< -3.75$                   \\
J1630$+$3734  &$1.013 \pm 0.015$           &$< -7.06$                   &$< -6.77$                   & $< -12.93$                  &--                          &$< -12.66$                  &--                          & $< -11.52$                  &--                          &$< -4.13$                   \\
J1640$+$2224  &$1.018 \pm 0.010$           &$-7.27 \pm 0.05$            &$-7.37 \,^{+0.11}_{-0.12}$  & $-13.15 \,^{+0.17}_{-0.18}$ &$0.83 \,^{+0.32}_{-0.33}$   &$-12.84 \,^{+0.19}_{-0.18}$ &$1.4 \,^{+0.5}_{-0.4}$      & $-12.08 \,^{+0.18}_{-0.21}$ &$0.62 \,^{+0.40}_{-0.33}$   &$< -4.12$                   \\
J1643$-$1224  &$1.14 \pm 0.07$             &$-6.328 \,^{+0.020}_{-0.023}$&$< -6.09$                  & $-12.36 \,^{+0.16}_{-0.14}$ &$1.9 \,^{+0.5}_{-0.4}$      &$-12.28 \,^{+0.17}_{-0.15}$ &$2.0 \pm 0.4$               & $-10.85 \,^{+0.15}_{-0.13}$ &$2.1 \pm 0.4$               &$-3.02 \,^{+0.12}_{-1.10}$  \\
J1710$+$4923  &$1.006 \pm 0.021$           &$< -6.64$                   &$< -6.44$                   & $< -12.84$                  &--                          &$< -12.42$                  &--                          & $< -11.39$                  &--                          &$< -3.58$                   \\
J1713$+$0747  &$0.831 \pm 0.006$           &$-7.852 \,^{+0.022}_{-0.023}$&$-7.171\,^{+0.024}_{-0.023}$&$-13.52 \,^{+0.31}_{-0.30}$ &$1.9 \,^{+0.6}_{-0.5}$       &$-13.6 \pm0.5$              &$1.7 \,^{+1.2}_{-0.7}$     & $< -12.04$                  &--                          &$< -4.53$                   \\
J1738$+$0333  &$1.030 \pm 0.008$           &$< -7.30$                   &$-7.02 \pm 0.06$            & $< -13.00$                  &--                          &$-13.41 \,^{+0.22}_{-0.19}$ &$0.6 \,^{+0.6}_{-0.4}$      & $-11.15 \,^{+0.11}_{-0.10}$ &$2.80 \,^{+0.40}_{-0.34}$   &$< -3.30$                   \\
J1741$+$1351  &$1.015 \pm 0.010$           &$< -7.43$                   &$-7.22 \pm 0.09$            & $< -13.40$                  &--                          &$< -13.35$                  &--                          & $-12.10 \,^{+0.13}_{-0.16}$ &$0.94 \,^{+0.50}_{-0.30}$   &$< -4.20$                   \\
J1744$-$1134  &$1.095 \pm 0.019$           &$-7.093 \,^{+0.019}_{-0.021}$&$-6.53 \pm 0.05$           & $< -13.23$                  &--                          &$< -13.22$                  &--                          & $-12.38 \pm 0.17$           &$0.47 \,^{+0.34}_{-0.29}$   &$< -4.55$                   \\
J1745$+$1017  &$1.081 \,^{+0.026}_{-0.027}$&$< -6.68$                   &$< -5.68$                   & $< -11.58$                  &--                          &$< -11.50$                  &--                          & $-11.02 \,^{+0.21}_{-0.20}$ &$0.9 \,^{+0.6}_{-0.5}$      &$< -2.93$                   \\
J1832$-$0836  &$0.968 \pm 0.011$           &$< -7.36$                   &$-6.99 \,^{+0.08}_{-0.09}$  & $< -13.08$                  &--                          &$< -13.03$                  &--                          & $-11.24 \,^{+0.11}_{-0.09}$ &$3.3 \,^{+0.5}_{-0.4}$      &$< -3.43$                   \\
J1843$-$1113  &$1.002 \pm 0.013$           &$< -7.09$                   &$< -6.96$                   & $< -13.08$                  &--                          &$< -12.90$                  &--                          & $-11.28 \,^{+0.11}_{-0.09}$ &$2.34 \,^{+0.32}_{-0.29}$   &$< -3.49$                   \\
J1853$+$1303  &$1.022 \pm 0.011$           &$< -7.30$                   &$< -7.11$                   & $< -13.18$                  &--                          &$< -13.05$                  &--                          & $< -11.82$                  &--                          &$< -4.20$                   \\
J1857$+$0943  &$0.968 \pm 0.015$           &$-7.25 \,^{+0.08}_{-0.12}$  &$-6.68 \pm 0.06$            & $< -13.28$                  &--                          &$< -13.05$                  &--                          & $-11.73 \pm 0.13$           &$2.1 \,^{+0.6}_{-0.5}$      &$< -3.93$                   \\
J1903$+$0327  &$1.143 \pm 0.014$           &$< -6.55$                   &$< -6.30$                   & $-12.03 \,^{+0.17}_{-0.14}$ &$3.8 \,^{+1.3}_{-1.0}$      &$-11.94 \,^{+0.18}_{-0.15}$ &$3.8 \,^{+1.0}_{-0.7}$      & $-10.52 \,^{+0.12}_{-0.11}$ &$2.8 \,^{+0.8}_{-0.4}$      &$< -2.73$                   \\
J1910$+$1256  &$1.006 \,^{+0.017}_{-0.016}$&$-6.88 \,^{+0.06}_{-0.07}$  &$-7.06 \,^{+0.09}_{-0.11}$  & $< -13.18$                  &--                          &$< -13.21$                  &--                          & $-11.65 \pm 0.13$           &$2.8 \,^{+0.8}_{-0.6}$      &$< -3.83$                   \\
J1911$-$1114  &$1.055 \pm 0.013$           &$< -6.66$                   &$< -6.64$                   & $< -12.84$                  &--                          &$< -12.47$                  &--                          & $-11.35 \,^{+0.14}_{-0.13}$ &$3.1 \,^{+0.8}_{-0.6}$      &$< -3.47$                   \\
J1911$+$1347  &$1.021 \,^{+0.014}_{-0.015}$&$< -7.20$                   &$-7.07 \pm 0.05$            & $< -13.82$                  &--                          &$< -13.53$                  &--                          & $-12.15 \,^{+0.16}_{-0.20}$ &$4.1 \,^{+1.4}_{-1.1}$      &$< -4.13$                   \\
J1918$-$0642  &$1.023 \pm 0.012$           &$< -7.23$                   &$< -6.82$                   & $< -12.99$                  &--                          &$< -12.80$                  &--                          & $-12.05 \,^{+0.21}_{-0.50}$ &$2.3 \,^{+3.2}_{-1.1}$      &$< -3.97$                   \\
J1923$+$2515  &$1.014 \pm 0.012$           &$< -7.08$                   &$< -6.79$                   & $< -13.07$                  &--                          &$< -12.96$                  &--                          & $< -11.61$                  &--                          &$< -3.80$                   \\
J1944$+$0907  &$1.024 \,^{+0.017}_{-0.016}$&$< -7.00$                   &$-6.39 \pm 0.07$            & $< -12.05$                  &--                          &$< -11.57$                  &--                          & $< -11.18$                  &--                          &$< -3.69$                   \\
J1946$+$3417  &$0.986 \,^{+0.011}_{-0.012}$&$< -6.60$                   &$-6.63 \pm 0.07$            & $-11.85 \,^{+0.18}_{-0.16}$ &$2.0 \pm 0.5$               &$-11.89 \,^{+0.15}_{-0.14}$ &$1.95 \,^{+0.40}_{-0.33}$   & $-10.15 \,^{+0.09}_{-0.10}$ &$2.41 \,^{+0.30}_{-0.34}$   &$< -2.49$                   \\
J1955$+$2908  &$1.002 \,^{+0.025}_{-0.022}$&$< -6.23$                   &$-6.64 \,^{+0.12}_{-0.15}$  & $< -12.48$                  &--                          &$< -12.20$                  &--                          & $-11.15 \pm 0.14$           &$2.5 \,^{+0.6}_{-0.4}$      &$< -3.35$                   \\
J2010$-$1323  &$1.068 \pm 0.017$           &$-6.76 \,^{+0.06}_{-0.07}$  &$< -6.66$                   & $< -12.61$                  &--                          &$< -12.87$                  &--                          & $< -11.50$                  &--                          &$< -4.00$                   \\
J2017$+$0603  &$1.009 \pm 0.012$           &$< -7.42$                   &$< -6.97$                   & $< -13.09$                  &--                          &$< -12.78$                  &--                          & $-11.87 \,^{+0.20}_{-0.28}$ &$4.7 \,^{+1.4}_{-1.3}$      &$< -3.83$                   \\
J2019$+$2425  &$1.070 \pm 0.012$           &$< -6.67$                   &$-6.72 \,^{+0.10}_{-0.12}$  & $< -13.08$                  &--                          &$< -11.93$                  &--                          & $< -11.62$                  &--                          &$< -3.77$                   \\
J2022$+$2534  &$1.053 \pm 0.020$           &$< -6.92$                   &$< -6.92$                   & $< -11.89$                  &--                          &$< -11.28$                  &--                          & $< -11.15$                  &--                          &$< -3.86$                   \\
J2033$+$1734  &$1.037 \,^{+0.015}_{-0.014}$&$-6.402 \,^{+0.034}_{-0.040}$&$< -6.56$                  & $< -13.20$                  &--                          &$< -12.58$                  &--                          & $< -11.68$                  &--                          &$-3.83 \,^{+0.13}_{-0.25}$  \\
J2043$+$1711  &$1.026 \,^{+0.015}_{-0.014}$&$-6.65 \pm 0.04$            &$< -7.02$                   & $< -12.95$                  &--                          &$-12.92 \,^{+0.25}_{-0.29}$ &$0.9 \,^{+0.6}_{-0.5}$      & $-11.77 \,^{+0.18}_{-0.20}$ &$3.9 \,^{+1.5}_{-1.1}$      &$< -3.72$                   \\
J2145$-$0750  &$0.863 \pm 0.011$           &$-7.26 \,^{+0.06}_{-0.07}$  &$-6.55 \pm 0.05$            & $< -13.14$                  &--                          &$< -12.70$                  &--                          & $< -11.56$                  &--                          &$< -4.02$                   \\
J2150$-$0326  &$1.001 \pm 0.019$           &$-7.14 \,^{+0.10}_{-0.16}$  &$-6.58 \,^{+0.10}_{-0.09}$  & $< -11.54$                  &--                          &$< -11.04$                  &--                          & $-11.34 \,^{+0.40}_{-0.33}$ &$2.5 \,^{+2.0}_{-1.5}$      &$< -2.80$                   \\
J2214$+$3000  &$1.051 \pm 0.013$           &$-6.622 \,^{+0.027}_{-0.028}$&$-6.73 \,^{+0.08}_{-0.09}$ & $< -12.89$                  &--                          &$< -12.38$                  &--                          & $-11.55 \,^{+0.14}_{-0.15}$ &$2.2 \,^{+1.2}_{-0.5}$      &$< -3.72$                   \\
J2229$+$2643  &$1.069 \,^{+0.013}_{-0.012}$&$-6.622 \,^{+0.029}_{-0.031}$&$< -6.67$                  & $< -12.99$                  &--                          &$-13.01 \,^{+0.18}_{-0.15}$ &$0.33 \,^{+0.40}_{-0.24}$   & $-12.15 \,^{+0.33}_{-0.40}$ &$5.1 \,^{+1.3}_{-1.9}$      &$< -3.89$                   \\
J2234$+$0611  &$1.060 \pm 0.012$           &$-7.15 \,^{+0.05}_{-0.06}$  &$-7.14 \pm 0.08$            & $< -13.39$                  &--                          &$< -12.77$                  &--                          & $< -11.96$                  &--                          &$< -4.14$                   \\
J2234$+$0944  &$0.988 \pm 0.011$           &$-7.19 \,^{+0.05}_{-0.06}$  &$-7.24 \,^{+0.19}_{-0.23}$  & $-12.96 \,^{+0.15}_{-0.14}$ &$1.71 \,^{+0.40}_{-0.35}$   &$-12.98 \,^{+0.20}_{-0.21}$ &$1.0 \pm 0.5$               & $-11.50 \,^{+0.14}_{-0.13}$ &$2.8 \,^{+0.9}_{-0.6}$      &$< -3.64$                   \\
J2302$+$4442  &$1.047 \,^{+0.011}_{-0.012}$&$< -6.76$                   &$-6.76 \,^{+0.07}_{-0.08}$  & $< -12.80$                  &--                          &$< -12.59$                  &--                          & $-11.56 \,^{+0.12}_{-0.10}$ &$1.67 \,^{+0.40}_{-0.29}$   &$< -3.81$                   \\
J2317$+$1439  &$1.047 \pm 0.011$           &$< -7.21$                   &$-6.83 \,^{+0.10}_{-0.15}$  & $< -13.19$                  &--                          &$< -12.66$                  &--                          & $< -11.58$                  &--                          &$< -3.92$                   \\
J2322$+$2057  &$1.027 \,^{+0.012}_{-0.011}$&$< -6.89$                   &$< -6.79$                   & $< -13.19$                  &--                          &$< -12.76$                  &--                          & $< -11.58$                  &--                          &$-3.88 \,^{+0.12}_{-0.19}$  \\
\hline\hline
\end{longtable}}
\tablefoot{Detections are given with $1\sigma$ uncertainties, while upper limits are set at 95\%.}
\end{center}

\setlength{\tabcolsep}{5.0pt}
\renewcommand{\arraystretch}{1.3}
\begin{center}
{\scriptsize
\begin{longtable}{c|ccc|cc|cc|ccc}
\caption{Summary of the noise parameters of the CPTA MSPs recovered using \texttt{42}.}
\label{tab:42}
\\\hline \hline
DM model & \multicolumn{3}{c|}{DM GP} & \multicolumn{2}{c|}{DM GP} & \multicolumn{2}{c|}{DMX} & \multicolumn{3}{c}{DM GP} \\
\hline
Pulsar name & EFAC & $\log_{10}$ EQUAD & $\log_{10}$ ECORR & $\log_{10}(A_{\rm RN})$ & $\gamma_{\rm RN}$ & $\log_{10}(A_{\rm RN})$ & $\gamma_{\rm RN}$ & $\log_{10}(A_{\rm DM})$ & $\gamma_{\rm DM}$ & $\log_{10}(A_{\rm yr})$ \\
\hline 
J0023$+$0923  &$1.047 \pm 0.006$            &$< -7.37$                    &$-6.95 \pm 0.06$              & $-13.28 \pm 0.17$           &$0.8 \pm 0.4$               &$-13.25 \,^{+0.28}_{-1.50}$ &$1.0 \,^{+2.5}_{-0.7}$      &$< -11.52$                   &--                           &$< -3.92$                    \\
J0030$+$0451  &$0.968 \pm 0.010$            &$< -7.26$                    &$-6.80 \,^{+0.09}_{-0.10}$    & $< -13.24$                  &--                          &$< -13.11$                  &--                          &$< -11.77$                   &--                           &$< -4.22$                    \\
J0034$-$0534  &$1.039 \,^{+0.016}_{-0.015}$ &$< -6.75$                    &$< -6.26$                     & $< -12.33$                  &--                          &$< -11.99$                  &--                          &$< -10.98$                   &--                           &$< -3.30$                    \\
J0154$+$1833  &$0.989 \pm 0.019$            &$< -7.11$                    &$-6.48 \,^{+0.11}_{-0.18}$    & $< -12.02$                  &--                          &$< -12.13$                  &--                          &$< -10.98$                   &--                           &$< -3.61$                    \\
J0218$+$4232  &$0.651 \,^{+0.006}_{-0.007}$ &$< -6.91$                    &$-6.83 \pm 0.04$              & $-13.11 \,^{+0.14}_{-0.15}$ &$1.3 \pm 0.4$               &$-12.91 \pm 0.14$           &$1.09 \,^{+0.40}_{-0.34}$   &$-11.18 \,^{+0.09}_{-0.08}$  &$2.27 \,^{+0.31}_{-0.25}$    &$< -3.51$                    \\
J0340$+$4130  &$1.019 \pm 0.017$            &$-6.50 \,^{+0.05}_{-0.06}$   &$< -7.03$                     & $< -13.30$                  &--                          &$-13.11 \,^{+0.20}_{-1.00}$ &$0.7 \,^{+1.2}_{-0.5}$      &$-11.54 \,^{+0.12}_{-0.11}$  &$1.86 \,^{+0.34}_{-0.31}$    &$-3.31 \,^{+0.05}_{-0.06}$   \\
J0406$+$3039  &$0.91 \pm 0.04$              &$-6.348 \,^{+0.033}_{-0.040}$&$< -6.66$                     & $< -11.37$                  &--                          &$< -11.15$                  &--                          &$-11.30 \,^{+0.26}_{-0.25}$  &$1.6 \,^{+0.7}_{-0.6}$       &$< -3.22$                    \\
J0509$+$0856  &$1.060 \,^{+0.020}_{-0.022}$ &$< -6.33$                    &$< -6.62$                     & $< -12.78$                  &--                          &$< -12.13$                  &--                          &$< -11.46$                   &--                           &$< -3.76$                    \\
J0605$+$3757  &$1.060 \pm 0.016$            &$< -6.57$                    &$< -6.27$                     & $< -12.08$                  &--                          &$< -11.83$                  &--                          &$< -10.94$                   &--                           &$< -3.32$                    \\
J0613$-$0200  &$0.957 \pm 0.010$            &$< -7.18$                    &$< -7.06$                     & $< -13.16$                  &--                          &$< -13.07$                  &--                          &$-11.71 \,^{+0.18}_{-0.23}$  &$3.6 \,^{+1.8}_{-1.2}$       &$< -3.65$                    \\
J0621$+$1002  &$0.939 \pm 0.010$            &$< -6.88$                    &$-5.84 \,^{+0.08}_{-0.18}$    & $< -12.21$                  &--                          &$< -12.24$                  &--                          &$-10.70 \,^{+0.09}_{-0.08}$  &$2.09 \,^{+0.25}_{-0.24}$    &$< -2.93$                    \\
J0636$+$5128  &$1.051 \pm 0.007$            &$-7.52 \,^{+0.05}_{-0.06}$   &$-6.829 \pm 0.019$            & $< -13.45$                  &--                          &$< -13.43$                  &--                          &$-12.38 \,^{+0.15}_{-0.16}$  &$0.74 \,^{+0.32}_{-0.31}$    &$< -4.40$                    \\
J0645$+$5158  &$1.047 \pm 0.011$            &$-7.57 \,^{+0.10}_{-0.13}$   &$-6.96 \,^{+0.07}_{-0.10}$    & $< -13.24$                  &--                          &$< -12.73$                  &--                          &$< -12.13$                   &--                           &$< -4.29$                    \\
J0732$+$2314  &$1.083 \pm 0.022$            &$< -6.58$                    &$< -6.27$                     & $< -12.27$                  &--                          &$< -11.63$                  &--                          &$< -11.20$                   &--                           &$< -3.70$                    \\
J0751$+$1807  &$1.001 \pm 0.006$            &$< -7.37$                    &$-7.01 \pm 0.04$              & $< -13.42$                  &--                          &$< -13.29$                  &--                          &$-11.78 \pm 0.12$            &$1.40 \,^{+0.40}_{-0.35}$    &$< -4.02$                    \\
J0824$+$0028  &$1.029 \,^{+0.022}_{-0.023}$ &$< -6.12$                    &$< -6.09$                     & $< -12.34$                  &--                          &$< -11.94$                  &--                          &$< -10.97$                   &--                           &$< -3.26$                    \\
J1012$+$5307  &$0.921 \pm 0.009$            &$< -7.46$                    &$-6.95 \pm 0.09$              & $-13.14 \,^{+0.19}_{-0.18}$ &$1.3 \pm 0.5$               &$-13.08 \,^{+0.19}_{-0.18}$ &$1.2 \pm 0.6$               &$< -11.87$                   &--                           &$< -4.15$                    \\
J1024$-$0719  &$1.024 \,^{+0.011}_{-0.012}$ &$< -7.04$                    &$< -7.00$                     & $< -12.97$                  &--                          &$< -12.13$                  &--                          &$-12.37 \,^{+0.20}_{-0.24}$  &$0.8 \,^{+0.5}_{-0.4}$       &$< -4.29$                    \\
J1327$+$3423  &$1.054 \,^{+0.019}_{-0.018}$ &$< -6.79$                    &$< -5.70$                     & $-12.48 \,^{+0.31}_{-0.19}$ &$0.6 \,^{+0.7}_{-0.4}$      &$-12.1 \,^{+0.5}_{-0.4}$    &$1.3 \,^{+1.4}_{-0.9}$      &$< -10.71$                   &--                           &$< -3.55$                    \\
J1453$+$1902  &$1.059 \pm 0.011$            &$< -6.58$                    &$< -6.53$                     & $< -12.78$                  &--                          &$< -12.65$                  &--                          &$< -11.44$                   &--                           &$< -3.75$                    \\
J1630$+$3734  &$1.013 \pm 0.015$            &$< -7.05$                    &$< -6.76$                     & $< -12.93$                  &--                          &$< -12.66$                  &--                          &$< -11.46$                   &--                           &$< -4.11$                    \\
J1640$+$2224  &$1.018 \,^{+0.011}_{-0.010}$ &$-7.26 \pm 0.05$             &$-7.35 \,^{+0.10}_{-0.11}$    & $-13.15 \,^{+0.17}_{-0.18}$ &$0.83 \,^{+0.32}_{-0.33}$   &$-12.84 \,^{+0.19}_{-0.18}$ &$1.4 \,^{+0.5}_{-0.4}$      &$-12.13 \,^{+0.19}_{-0.21}$  &$0.50 \,^{+0.33}_{-0.30}$    &$< -4.17$                    \\
J1643$-$1224  &$1.14 \pm 0.07$              &$-6.327 \,^{+0.021}_{-0.024}$&$< -6.21$                     & $-12.36 \,^{+0.16}_{-0.14}$ &$1.9 \,^{+0.5}_{-0.4}$      &$-12.28 \,^{+0.17}_{-0.15}$ &$2.0 \pm 0.4$               &$-10.79 \,^{+0.15}_{-0.16}$  &$2.3 \,^{+0.5}_{-0.4}$       &$< -2.82$                    \\
J1710$+$4923  &$1.005 \,^{+0.022}_{-0.021}$ &$< -6.63$                    &$< -6.44$                     & $< -12.84$                  &--                          &$< -12.42$                  &--                          &$< -11.43$                   &--                           &$< -3.59$                    \\
J1713$+$0747  &$0.831 \pm 0.006$           &$-7.851 \,^{+0.022}_{-0.023}$&$-7.171 \pm 0.024$            & $-13.52 \,^{+0.31}_{-0.30}$ &$1.9 \,^{+0.6}_{-0.5}$      &$-13.6 \pm0.5$              &$1.7 \,^{+1.2}_{-0.7}$      &$< -11.95$                   &--                           &$< -4.54$                    \\
J1738$+$0333  &$1.030 \pm 0.008$            &$< -7.29$                    &$-7.01 \pm 0.06$              & $< -13.00$                  &--                          &$-13.41 \,^{+0.22}_{-0.19}$ &$0.6 \,^{+0.6}_{-0.4}$      &$-11.14 \pm 0.10$            &$2.84 \,^{+0.40}_{-0.34}$    &$< -3.30$                    \\
J1741$+$1351  &$1.015 \pm 0.010$            &$< -7.44$                    &$-7.19 \pm 0.10$              & $< -13.40$                  &--                          &$< -13.35$                  &--                          &$-12.08 \,^{+0.12}_{-0.14}$  &$0.95 \,^{+0.40}_{-0.29}$    &$< -4.22$                    \\
J1744$-$1134  &$1.096 \pm 0.019$            &$-7.080 \,^{+0.019}_{-0.020}$&$-6.54 \pm 0.05$              & $< -13.23$                  &--                          &$< -13.22$                  &--                          &$-12.25 \,^{+0.23}_{-0.33}$  &$3.0 \,^{+2.1}_{-1.2}$       &$< -4.34$                    \\
J1745$+$1017  &$1.081 \,^{+0.025}_{-0.026}$ &$< -6.67$                    &$< -5.68$                     & $< -11.58$                  &--                          &$< -11.50$                  &--                          &$-10.94 \,^{+0.19}_{-0.20}$  &$1.1 \pm 0.5$                &$< -2.93$                    \\
J1832$-$0836  &$0.968 \pm 0.011$            &$< -7.36$                    &$-6.93 \pm 0.07$              & $< -13.08$                  &--                          &$< -13.03$                  &--                          &$-11.19 \,^{+0.12}_{-0.10}$  &$3.9 \pm 0.5$                &$< -3.40$                    \\
J1843$-$1113  &$1.002 \pm 0.013$            &$< -7.09$                    &$< -6.95$                     & $< -13.08$                  &--                          &$< -12.90$                  &--                          &$-11.26 \pm 0.10$            &$2.42 \,^{+0.33}_{-0.30}$    &$< -3.45$                    \\
J1853$+$1303  &$1.022 \pm 0.011$            &$< -7.30$                    &$< -7.11$                     & $< -13.18$                  &--                          &$< -13.05$                  &--                          &$< -11.78$                   &--                           &$< -4.21$                    \\
J1857$+$0943  &$0.966 \pm 0.015$            &$-7.25 \,^{+0.08}_{-0.12}$   &$-6.70 \pm 0.07$              & $< -13.28$                  &--                          &$< -13.05$                  &--                          &$-11.66 \pm 0.12$            &$2.0 \,^{+0.5}_{-0.4}$       &$< -3.87$                    \\
J1903$+$0327  &$1.143 \pm 0.014$            &$< -6.56$                    &$< -6.25$                     & $-12.03 \,^{+0.17}_{-0.14}$ &$3.8 \,^{+1.3}_{-1.0}$      &$-11.94 \,^{+0.18}_{-0.15}$ &$3.8 \,^{+1.0}_{-0.7}$      &$-10.50 \,^{+0.13}_{-0.11}$  &$3.5 \,^{+0.9}_{-1.2}$       &$< -2.76$                    \\
J1910$+$1256  &$1.006 \pm 0.016$            &$-6.88 \,^{+0.06}_{-0.07}$   &$< -6.90$                     & $< -13.18$                  &--                          &$< -13.21$                  &--                          &$-11.67 \pm 0.13$            &$2.5 \,^{+0.9}_{-0.6}$       &$< -3.81$                    \\
J1911$-$1114  &$1.055 \,^{+0.013}_{-0.012}$ &$< -6.65$                    &$< -6.63$                     & $< -12.84$                  &--                          &$< -12.47$                  &--                          &$-11.32 \,^{+0.15}_{-0.14}$  &$3.4 \,^{+0.9}_{-0.7}$       &$< -3.48$                    \\
J1911$+$1347  &$1.021 \,^{+0.014}_{-0.015}$ &$< -7.21$                    &$-7.06 \pm 0.05$              & $< -13.82$                  &--                          &$< -13.53$                  &--                          &$-12.09 \,^{+0.15}_{-0.20}$  &$4.2 \,^{+1.4}_{-1.2}$       &$< -4.11$                    \\
J1918$-$0642  &$1.023 \pm 0.012$            &$< -7.24$                    &$-7.05 \pm 0.14$              & $< -12.99$                  &--                          &$< -12.80$                  &--                          &$-11.94 \,^{+0.17}_{-0.29}$  &$2.2 \,^{+3.0}_{-0.9}$       &$< -3.97$                    \\
J1923$+$2515  &$1.014 \,^{+0.012}_{-0.011}$ &$< -7.08$                    &$< -6.79$                     & $< -13.07$                  &--                          &$< -12.96$                  &--                          &$< -11.58$                   &--                           &$< -3.80$                    \\
J1944$+$0907  &$1.024 \,^{+0.017}_{-0.016}$ &$< -6.98$                    &$-6.39 \pm 0.07$              & $< -12.05$                  &--                          &$< -11.57$                  &--                          &$< -11.11$                   &--                           &$< -3.69$                    \\
J1946$+$3417  &$0.986 \,^{+0.011}_{-0.012}$ &$< -6.61$                    &$-6.63 \pm 0.07$              & $-11.85 \,^{+0.18}_{-0.16}$ &$2.0 \pm 0.5$               &$-11.89 \,^{+0.15}_{-0.14}$ &$1.95 \,^{+0.40}_{-0.33}$   &$-10.11 \,^{+0.07}_{-0.09}$  &$2.52 \,^{+0.27}_{-0.32}$    &$< -2.49$                    \\
J1955$+$2908  &$1.001 \,^{+0.026}_{-0.022}$&$< -6.23$                     &$< -6.50$                     & $< -12.48$                  &--                          &$< -12.20$                  &--                          &$-11.14 \pm 0.13$            &$2.3 \,^{+0.7}_{-0.4}$       &$< -3.32$                    \\
J2010$-$1323  &$1.068 \pm 0.017$            &$-6.76 \,^{+0.06}_{-0.07}$   &$< -6.66$                     & $< -12.61$                  &--                          &$< -12.87$                  &--                          &$< -11.49$                   &--                           &$< -4.02$                    \\
J2017$+$0603  &$1.009 \pm 0.012$            &$< -7.41$                    &$< -6.97$                     & $< -13.09$                  &--                          &$< -12.78$                  &--                          &$-11.81 \,^{+0.17}_{-0.20}$  &$5.1 \pm 1.3$                &$< -3.83$                    \\
J2019$+$2425  &$1.070 \pm 0.012$            &$< -6.66$                    &$-6.72 \,^{+0.10}_{-0.12}$    & $< -13.08$                  &--                          &$< -11.93$                  &--                          &$< -11.55$                   &--                           &$< -3.75$                    \\
J2022$+$2534  &$1.053 \pm 0.020$            &$< -6.92$                    &$< -6.91$                     & $< -11.89$                  &--                          &$< -11.28$                  &--                          &$< -11.01$                   &--                           &$< -3.85$                    \\
J2033$+$1734  &$1.037 \pm 0.015$            &$-6.401 \,^{+0.035}_{-0.040}$&$< -6.55$                     & $< -13.20$                  &--                          &$< -12.58$                  &--                          &$< -11.69$                   &--                           &$-3.81 \,^{+0.12}_{-0.20}$   \\
J2043$+$1711  &$1.026 \pm 0.015$            &$-6.65 \pm 0.04$             &$< -7.05$                     & $< -12.95$                  &--                          &$-12.92 \,^{+0.25}_{-0.29}$ &$0.9 \,^{+0.6}_{-0.5}$      &$-11.72 \,^{+0.16}_{-0.18}$  &$4.3 \,^{+1.4}_{-1.3}$       &$< -3.71$                    \\
J2145$-$0750  &$0.863 \,^{+0.012}_{-0.011}$ &$-7.26 \,^{+0.06}_{-0.07}$   &$-6.55 \pm 0.05$              & $< -13.14$                  &--                          &$< -12.70$                  &--                          &$< -11.50$                   &--                           &$< -4.01$                    \\
J2150$-$0326  &$1.002 \,^{+0.020}_{-0.019}$ &$-7.14 \,^{+0.10}_{-0.17}$   &$-6.58 \,^{+0.10}_{-0.09}$    & $< -11.54$                  &--                          &$< -11.04$                  &--                          &$-11.26 \,^{+0.40}_{-0.35}$  &$2.9 \,^{+2.1}_{-1.6}$       &$< -3.06$                    \\
J2214$+$3000  &$1.051 \pm 0.013$            &$-6.617 \,^{+0.027}_{-0.028}$&$-6.69 \,^{+0.06}_{-0.07}$    & $< -12.89$                  &--                          &$< -12.38$                  &--                          &$-11.59 \,^{+0.17}_{-0.25}$  &$3.8 \,^{+1.9}_{-1.3}$       &$< -3.72$                    \\
J2229$+$2643  &$1.069 \pm 0.013$            &$-6.622 \,^{+0.029}_{-0.030}$&$< -6.96$                     & $< -12.99$                  &--                          &$-13.01 \,^{+0.18}_{-0.15}$ &$0.33 \,^{+0.40}_{-0.24}$   &$-12.05 \,^{+0.26}_{-0.27}$  &$5.6 \,^{+1.0}_{-1.6}$       &$< -3.91$                    \\
J2234$+$0611  &$1.060 \pm 0.012$            &$-7.15 \,^{+0.05}_{-0.06}$   &$-7.13 \,^{+0.07}_{-0.08}$    & $< -13.39$                  &--                          &$< -12.77$                  &--                          &$-12.44 \,^{+0.27}_{-0.30}$  &$5.2 \,^{+1.2}_{-1.7}$       &$< -4.13$                    \\
J2234$+$0944  &$0.988 \pm 0.011$            &$-7.18 \,^{+0.05}_{-0.06}$   &$< -6.93$                     & $-12.96 \,^{+0.15}_{-0.14}$ &$1.71 \,^{+0.40}_{-0.35}$   &$-12.98 \,^{+0.20}_{-0.21}$ &$1.0 \pm 0.5$               &$-11.44 \,^{+0.14}_{-0.13}$  &$3.4 \,^{+1.0}_{-0.7}$       &$< -3.58$                    \\
J2302$+$4442  &$1.048 \,^{+0.011}_{-0.012}$ &$< -6.76$                    &$-6.63 \pm 0.06$              & $< -12.80$                  &--                          &$< -12.59$                  &--                          &$-11.51 \,^{+0.14}_{-0.15}$  &$2.9 \,^{+1.3}_{-0.7}$       &$< -3.73$                    \\
J2317$+$1439  &$1.046 \pm 0.011$            &$< -7.22$                    &$-6.85 \,^{+0.10}_{-0.14}$    & $< -13.19$                  &--                          &$< -12.66$                  &--                          &$< -11.58$                   &--                           &$< -3.93$                    \\
J2322$+$2057  &$1.028 \,^{+0.011}_{-0.012}$ &$< -6.89$                    &$< -6.77$                     & $< -13.19$                  &--                          &$< -12.76$                  &--                          &$< -11.54$                   &--                           &$-3.88 \,^{+0.12}_{-0.18}$   \\
\hline\hline
\end{longtable}}
\tablefoot{Detections are given with $1\sigma$ uncertainties, while upper limits are set at 95\%.}
\end{center}

\setlength{\tabcolsep}{5.0pt}
\renewcommand{\arraystretch}{1.3}
\begin{center}
{\scriptsize
\begin{longtable}{c|ccc|cc|cc|ccc}
\caption{Summary of the noise parameters of the CPTA MSPs recovered using \texttt{42++}.}
\label{tab:42pp}
\\\hline \hline
DM model & \multicolumn{3}{c|}{DM GP} & \multicolumn{2}{c|}{DM GP} & \multicolumn{2}{c|}{DMX} & \multicolumn{3}{c}{DM GP} \\
\hline
Pulsar name & EFAC & $\log_{10}$ EQUAD & $\log_{10}$ ECORR & $\log_{10}(A_{\rm RN})$ & $\gamma_{\rm RN}$ & $\log_{10}(A_{\rm RN})$ & $\gamma_{\rm RN}$ & $\log_{10}(A_{\rm DM})$ & $\gamma_{\rm DM}$ & $\log_{10}(A_{\rm yr})$ \\
\hline 
J0023$+$0923  &$1.046 \pm 0.006$           &$< -7.38$                   &$-6.95 \,^{+0.05}_{-0.06}$  & $-13.19 \,^{+0.07}_{-0.13}$ &$0.50 \,^{+0.50}_{-0.34}$   &$< -12.64$                  &--                          & $< -11.58$                  &--                          &$-4.15 \,^{+0.13}_{-0.26}$  \\
J0030$+$0451  &$0.968 \pm 0.009$           &$< -7.25$                   &$-6.97 \pm 0.09$            & $< -13.31$                  &--                          &$< -12.85$                  &--                          & $-12.04 \,^{+0.09}_{-0.19}$ &$0.6 \pm 0.4$               &$< -4.24$                   \\
J0034$-$0534  &$1.035 \,^{+0.014}_{-0.012}$&$< -6.81$                   &$< -6.36$                   & $< -12.34$                  &--                          &$< -12.10$                  &--                          & $< -11.01$                  &--                          &$< -3.30$                   \\
J0154$+$1833  &$0.988 \,^{+0.017}_{-0.016}$&$< -7.16$                   &$< -6.30$                   & $< -12.13$                  &--                          &$< -12.05$                  &--                          & $< -11.05$                  &--                          &$< -3.62$                   \\
J0218$+$4232  &$0.651 \pm 0.006$           &$< -6.94$                   &$-6.91 \,^{+0.04}_{-0.05}$  & $-13.06 \,^{+0.14}_{-0.20}$ &$1.7 \pm 0.4$               &$-12.95 \,^{+0.08}_{-0.17}$ &$0.71 \,^{+0.32}_{-0.33}$   & $-11.24 \,^{+0.08}_{-0.07}$ &$2.18 \,^{+0.25}_{-0.21}$   &$< -3.56$                   \\
J0340$+$4130  &$1.018 \pm 0.015$           &$-6.50 \,^{+0.04}_{-0.05}$  &$< -7.08$                   & $< -13.36$                  &--                          &$< -12.63$                  &--                          & $-11.51 \,^{+0.12}_{-0.11}$ &$1.98 \,^{+0.32}_{-0.29}$   &$-3.32 \,^{+0.06}_{-0.07}$  \\
J0406$+$3039  &$0.91 \pm 0.04$             &$-6.346 \,^{+0.032}_{-0.035}$&$< -6.68$                  & $< -11.42$                  &--                          &$< -11.39$                  &--                          & $-11.23 \,^{+0.25}_{-0.30}$ &$1.9 \pm 0.6$               &$< -3.18$                   \\
J0509$+$0856  &$1.065 \,^{+0.015}_{-0.016}$&$< -6.40$                   &$< -6.64$                   & $< -12.61$                  &--                          &$< -12.08$                  &--                          & $< -11.39$                  &--                          &$< -3.81$                   \\
J0605$+$3757  &$1.060 \,^{+0.014}_{-0.015}$&$< -6.56$                   &$< -6.26$                   & $< -12.24$                  &--                          &$< -11.98$                  &--                          & $< -11.07$                  &--                          &$< -3.35$                   \\
J0613$-$0200  &$0.958 \pm 0.009$           &$< -7.21$                   &$-7.25 \,^{+0.16}_{-0.15}$  & $< -13.22$                  &--                          &$< -13.05$                  &--                          & $-11.79 \,^{+0.18}_{-0.40}$ &$4.1 \,^{+2.5}_{-1.6}$      &$-3.84 \,^{+0.09}_{-0.19}$  \\
J0621$+$1002  &$0.939 \pm 0.009$           &$< -6.90$                   &$-5.86 \,^{+0.09}_{-0.17}$  & $< -12.22$                  &--                          &$< -12.26$                  &--                          & $-10.71 \,^{+0.09}_{-0.08}$ &$2.11 \pm 0.22$             &$< -2.98$                   \\
J0636$+$5128  &$1.050 \pm 0.006$           &$-7.52 \,^{+0.04}_{-0.05}$  &$-6.832 \pm 0.017$          & $< -13.63$                  &--                          &$-12.68 \pm 0.05$           &$0.23 \,^{+0.26}_{-0.15}$   & $-12.24 \,^{+0.07}_{-0.08}$ &$0.35 \,^{+0.35}_{-0.24}$   &$< -4.45$                   \\
J0645$+$5158  &$1.047 \pm 0.009$           &$-7.58 \,^{+0.09}_{-0.13}$  &$< -6.84$                   & $< -13.36$                  &--                          &$< -12.73$                  &--                          & $< -12.19$                  &--                          &$< -4.32$                   \\
J0732$+$2314  &$1.084 \,^{+0.020}_{-0.018}$&$< -6.57$                   &$< -6.30$                   & $< -12.17$                  &--                          &$< -11.83$                  &--                          & $< -11.28$                  &--                          &$< -3.71$                   \\
J0751$+$1807  &$1.001 \pm 0.006$           &$< -7.39$                   &$-7.03 \pm 0.04$            & $< -13.44$                  &--                          &$< -13.27$                  &--                          & $-11.90 \,^{+0.16}_{-0.23}$ &$1.33 \,^{+0.30}_{-0.31}$   &$< -4.09$                   \\
J0824$+$0028  &$1.034 \pm 0.019$           &$< -6.14$                   &$< -6.13$                   & $< -12.26$                  &--                          &$< -12.13$                  &--                          & $< -10.98$                  &--                          &$< -3.26$                   \\
J1012$+$5307  &$0.920 \,^{+0.009}_{-0.008}$&$< -7.48$                   &$-7.09 \,^{+0.10}_{-0.09}$  & $-13.14 \,^{+0.19}_{-0.25}$ &$1.5 \pm 0.4$               &$-13.09 \pm 0.20$           &$1.1 \,^{+0.5}_{-0.6}$      & $< -11.98$                  &--                          &$< -4.13$                   \\
J1024$-$0719  &$1.025 \,^{+0.010}_{-0.009}$&$< -7.07$                   &$< -7.12$                   & $< -13.05$                  &--                          &$< -12.18$                  &--                          & $-12.27 \,^{+0.11}_{-0.19}$ &$0.7 \,^{+0.5}_{-0.4}$      &$< -4.25$                   \\
J1327$+$3423  &$1.055 \,^{+0.016}_{-0.015}$&$< -6.81$                   &$< -5.65$                   & $-12.42 \,^{+0.17}_{-1.40}$ &$0.9 \,^{+2.7}_{-0.6}$      &$-12.1 \pm 0.4$             &$1.7 \,^{+1.1}_{-1.0}$      & $< -10.82$                  &--                          &$< -3.62$                   \\
J1453$+$1902  &$1.060 \,^{+0.010}_{-0.009}$&$< -6.61$                   &$< -6.54$                   & $< -12.66$                  &--                          &$< -12.70$                  &--                          & $< -11.55$                  &--                          &$< -3.84$                   \\
J1630$+$3734  &$1.014 \pm 0.013$           &$< -7.09$                   &$< -6.79$                   & $< -12.88$                  &--                          &$< -12.68$                  &--                          & $< -11.46$                  &--                          &$< -4.06$                   \\
J1640$+$2224  &$1.017 \pm 0.009$           &$-7.26 \pm 0.04$            &$-7.38 \,^{+0.10}_{-0.11}$  & $-13.12 \pm 0.05$           &$0.21 \,^{+0.35}_{-0.15}$   &$-13.01 \,^{+0.22}_{-0.20}$ &$1.1 \,^{+0.4}_{-0.5}$      & $-11.90 \,^{+0.09}_{-0.10}$ &$0.30 \,^{+0.29}_{-0.20}$   &$< -4.24$                   \\
J1643$-$1224  &$1.13 \pm 0.07$             &$-6.326 \,^{+0.019}_{-0.023}$&$< -6.13$                  & $-12.38 \,^{+0.15}_{-0.14}$ &$1.88 \,^{+0.35}_{-0.31}$   &$-12.31 \,^{+0.14}_{-0.15}$ &$2.02 \,^{+0.40}_{-0.35}$   & $-10.86 \,^{+0.15}_{-0.13}$ &$2.16 \,^{+0.40}_{-0.34}$   &$-3.04 \,^{+0.13}_{-0.90}$  \\
J1710$+$4923  &$1.006 \,^{+0.019}_{-0.017}$&$< -6.69$                   &$< -6.46$                   & $< -12.74$                  &--                          &$< -12.16$                  &--                          & $< -11.47$                  &--                          &$< -3.61$                   \\
J1713$+$0747  &$0.830 \pm 0.006$           &$-7.852 \,^{+0.020}_{-0.021}$&$-7.171 \,^{+0.022}_{-0.025}$ & $-13.66 \,^{+0.34}_{-0.40}$ &$1.7 \pm 0.7$               &                            &                           & $< -12.07$                  &--                          &$-4.73 \,^{+0.12}_{-0.26}$  \\
J1738$+$0333  &$1.030 \pm 0.007$           &$< -7.30$                   &$-7.02 \pm 0.06$            & $< -12.99$                  &--                          &$-13.26 \,^{+0.10}_{-0.16}$ &$0.53 \,^{+0.50}_{-0.34}$   & $-11.16 \pm 0.10$           &$2.78 \,^{+0.34}_{-0.33}$   &$< -3.31$                   \\
J1741$+$1351  &$1.017 \pm 0.009$           &$< -7.44$                   &$-7.21 \,^{+0.09}_{-0.08}$  & $< -13.42$                  &--                          &$< -13.33$                  &--                          & $-12.14 \,^{+0.14}_{-0.24}$ &$1.00 \pm 0.26$             &$< -4.19$                   \\
J1744$-$1134  &$1.094 \,^{+0.018}_{-0.017}$&$-7.094 \,^{+0.018}_{-0.020}$&$-6.53 \pm 0.05$           & $< -13.14$                  &--                          &$< -13.14$                  &--                          & $-12.10 \,^{+0.09}_{-0.15}$ &$0.44 \,^{+0.40}_{-0.30}$   &$< -4.53$                   \\
J1745$+$1017  &$1.085 \pm 0.022$           &$< -6.70$                   &$< -5.71$                   & $< -11.87$                  &--                          &$< -11.89$                  &--                          & $-11.00 \,^{+0.17}_{-0.20}$ &$0.9 \pm 0.5$               &$< -2.99$                   \\
J1832$-$0836  &$0.969 \pm 0.010$           &$< -7.39$                   &$< -6.97$                   & $< -13.04$                  &--                          &$< -12.98$                  &--                          & $-11.22 \pm 0.08$           &$2.86 \,^{+0.40}_{-0.26}$   &$< -3.49$                   \\
J1843$-$1113  &$1.002 \pm 0.011$           &$< -7.11$                   &$< -7.04$                   & $< -13.10$                  &--                          &$< -12.67$                  &--                          & $-11.27 \,^{+0.10}_{-0.09}$ &$2.36 \,^{+0.26}_{-0.24}$   &$< -3.44$                   \\
J1853$+$1303  &$1.022 \pm 0.010$           &$< -7.30$                   &$< -7.13$                   & $< -13.13$                  &--                          &$< -12.86$                  &--                          & $< -11.78$                  &--                          &$< -4.20$                   \\
J1857$+$0943  &$0.966 \,^{+0.015}_{-0.013}$&$-7.25 \,^{+0.08}_{-0.12}$  &$-6.73 \,^{+0.07}_{-0.08}$  & $< -13.31$                  &--                          &$< -13.01$                  &--                          & $-11.74 \,^{+0.12}_{-0.15}$ &$1.67 \,^{+0.40}_{-0.33}$   &$< -3.92$                   \\
J1903$+$0327  &$1.143 \pm 0.013$           &$< -6.58$                   &$< -6.33$                   & $-12.02 \,^{+0.08}_{-0.10}$ &$4.1 \,^{+0.6}_{-0.7}$      &$-11.95 \,^{+0.08}_{-0.10}$ &$3.7 \pm 0.5$               & $-10.55 \,^{+0.10}_{-0.09}$ &$2.62 \,^{+0.35}_{-0.27}$   &$< -2.90$                   \\
J1910$+$1256  &$1.006 \,^{+0.015}_{-0.014}$&$-6.88 \,^{+0.05}_{-0.06}$  &$< -6.95$                   & $< -13.16$                  &--                          &$< -13.18$                  &--                          & $-11.73 \,^{+0.11}_{-0.12}$ &$1.84 \,^{+0.34}_{-0.32}$   &$< -3.92$                   \\
J1911$-$1114  &$1.056 \pm 0.011$           &$< -6.68$                   &$< -6.65$                   & $< -12.87$                  &--                          &$< -12.43$                  &--                          & $-11.32 \pm 0.12$           &$3.0 \,^{+0.7}_{-0.6}$      &$< -3.56$                   \\
J1911$+$1347  &$1.024 \,^{+0.012}_{-0.014}$&$< -7.22$                   &$-7.06 \,^{+0.04}_{-0.05}$  & $< -13.73$                  &--                          &$< -13.50$                  &--                          & $-12.12 \,^{+0.17}_{-0.20}$ &$4.1 \,^{+1.6}_{-1.4}$      &$< -4.13$                   \\
J1918$-$0642  &$1.024 \,^{+0.012}_{-0.011}$&$< -7.28$                   &$-6.99 \,^{+0.14}_{-0.16}$  & $< -12.97$                  &--                          &$< -12.66$                  &--                          & $-12.19 \,^{+0.32}_{-0.50}$ &$4.6 \,^{+3.1}_{-2.9}$      &$< -4.00$                   \\
J1923$+$2515  &$1.015 \pm 0.010$           &$< -7.09$                   &$< -6.81$                   & $< -13.01$                  &--                          &$< -12.82$                  &--                          & $< -11.55$                  &--                          &$< -3.80$                   \\
J1944$+$0907  &$1.023 \pm 0.015$           &$< -7.01$                   &$-6.39 \,^{+0.07}_{-0.06}$  & $< -11.85$                  &--                          &$< -11.80$                  &--                          & $< -11.08$                  &--                          &$< -3.69$                   \\
J1946$+$3417  &$0.987 \pm 0.010$           &$< -6.69$                   &$-6.63 \pm 0.07$            & $-11.83 \,^{+0.11}_{-0.15}$ &$2.06 \,^{+0.33}_{-0.40}$   &$-11.83 \,^{+0.10}_{-0.12}$ &$2.12 \,^{+0.25}_{-0.32}$   & $-10.07 \,^{+0.13}_{-0.12}$ &$2.7 \pm 0.4$               &$< -2.46$                   \\
J1955$+$2908  &$1.007 \,^{+0.027}_{-0.021}$&$-6.41 \,^{+0.10}_{-1.60}$   &$< -6.51$                   & $< -12.54$                  &--                          &$< -12.24$                  &--                          & $-11.14 \pm 0.10$           &$2.20 \,^{+0.40}_{-0.32}$   &$< -3.37$                   \\
J2010$-$1323  &$1.069 \,^{+0.015}_{-0.014}$&$-6.77 \,^{+0.05}_{-0.06}$  &$< -6.66$                   & $-13.34 \,^{+0.22}_{-0.31}$ &$1.3 \pm 0.5$               &$< -12.89$                  &--                          & $< -11.54$                  &--                          &$< -4.02$                   \\
J2017$+$0603  &$1.008 \,^{+0.011}_{-0.010}$&$< -7.46$                   &$< -6.98$                   & $< -13.02$                  &--                          &$< -12.83$                  &--                          & $-11.85 \,^{+0.18}_{-0.25}$ &$5.5 \,^{+1.9}_{-1.7}$      &$< -3.86$                   \\
J2019$+$2425  &$1.071 \,^{+0.010}_{-0.011}$&$< -6.70$                   &$-6.73 \,^{+0.09}_{-0.12}$  & $< -12.94$                  &--                          &$< -12.02$                  &--                          & $< -11.59$                  &--                          &$< -3.81$                   \\
J2022$+$2534  &$1.055 \pm 0.017$           &$< -6.92$                   &$< -6.92$                   & $< -11.58$                  &--                          &$< -11.41$                  &--                          & $< -10.73$                  &--                          &$< -3.78$                   \\
J2033$+$1734  &$1.040 \pm 0.013$           &$-6.416 \,^{+0.033}_{-0.035}$&$< -6.59$                  & $< -13.14$                  &--                          &$< -12.64$                  &--                          & $-11.88 \,^{+0.07}_{-0.09}$ &$0.31 \,^{+0.35}_{-0.21}$   &$-3.82 \,^{+0.11}_{-0.18}$  \\
J2043$+$1711  &$1.026 \pm 0.014$           &$-6.651 \,^{+0.035}_{-0.040}$&$< -7.04$                  & $< -12.75$                  &--                          &$-12.73 \,^{+0.10}_{-0.16}$ &$0.48 \,^{+0.50}_{-0.31}$   & $-11.71 \,^{+0.16}_{-0.18}$ &$4.2 \,^{+1.6}_{-1.3}$      &$< -3.72$                   \\
J2145$-$0750  &$0.862 \,^{+0.011}_{-0.010}$&$-7.26 \,^{+0.06}_{-0.07}$  &$-6.55 \pm 0.05$            & $< -12.94$                  &--                          &$< -12.68$                  &--                          & $< -11.55$                  &--                          &$< -4.02$                   \\
J2150$-$0326  &$1.002 \pm 0.018$           &$-7.14 \,^{+0.10}_{-0.18}$  &$-6.58 \pm 0.09$            & $< -11.50$                  &--                          &$< -11.40$                  &--                          & $-11.2 \pm 0.4$             &$3.3 \,^{+2.2}_{-1.7}$      &$< -2.93$                   \\
J2214$+$3000  &$1.051 \,^{+0.013}_{-0.012}$&$-6.622 \,^{+0.026}_{-0.028}$&$-6.73 \,^{+0.08}_{-0.10}$ & $< -12.72$                  &--                          &$-12.84 \,^{+0.09}_{-0.19}$ &$0.40 \,^{+0.60}_{-0.27}$   & $-11.52 \,^{+0.13}_{-0.19}$ &$2.3 \,^{+2.5}_{-0.5}$      &$< -3.74$                   \\
J2229$+$2643  &$1.068 \pm 0.012$           &$-6.621 \,^{+0.027}_{-0.029}$&$< -6.68$                  & $-13.48 \,^{+0.12}_{-0.19}$ &$0.7 \,^{+0.6}_{-0.5}$      &$-12.73 \,^{+0.09}_{-0.14}$ &$0.53 \,^{+0.40}_{-0.33}$   & $-12.22 \,^{+0.35}_{-0.32}$ &$6.8 \,^{+1.5}_{-2.0}$      &$< -3.90$                   \\
J2234$+$0611  &$1.059 \pm 0.011$           &$-7.15 \pm 0.05$            &$-7.12 \pm 0.07$            & $< -13.46$                  &--                          &$< -12.91$                  &--                          & $-12.62 \,^{+0.33}_{-0.34}$ &$6.5 \,^{+1.6}_{-1.9}$      &$< -4.15$                   \\
J2234$+$0944  &$0.987 \pm 0.010$           &$-7.18 \,^{+0.05}_{-0.06}$  &$< -6.99$                   & $-12.98 \pm 0.14$           &$1.76 \pm 0.30$             &$-12.97 \,^{+0.09}_{-0.14}$ &$0.53 \,^{+0.40}_{-0.33}$   & $-11.45 \,^{+0.13}_{-0.12}$ &$3.4 \,^{+0.9}_{-0.7}$      &$< -3.59$                   \\
J2302$+$4442  &$1.049 \pm 0.010$           &$< -6.78$                   &$-6.71 \,^{+0.10}_{-0.11}$  & $< -12.70$                  &--                          &$< -12.64$                  &--                          & $-11.51 \pm 0.12$           &$2.1 \,^{+1.4}_{-0.4}$      &$< -3.79$                   \\
J2317$+$1439  &$1.047 \pm 0.010$           &$< -7.23$                   &$-6.98 \,^{+0.15}_{-0.17}$  & $< -13.15$                  &--                          &$< -12.62$                  &--                          & $-12.03 \,^{+0.15}_{-0.22}$ &$0.9 \,^{+0.8}_{-0.6}$      &$< -3.92$                   \\
J2322$+$2057  &$1.028 \,^{+0.011}_{-0.010}$&$< -6.90$                   &$< -6.79$                   & $< -13.37$                  &--                          &$< -12.51$                  &--                          & $< -11.69$                  &--                          &$-3.88 \,^{+0.12}_{-0.18}$  \\
\hline\hline
\end{longtable}}
\tablefoot{Detections are given with $1\sigma$ uncertainties, while upper limits are set at 95\%.}
\end{center}

\section{Comparison with results from other PTAs}
\begin{figure*}
\centering
\includegraphics[width=\linewidth]{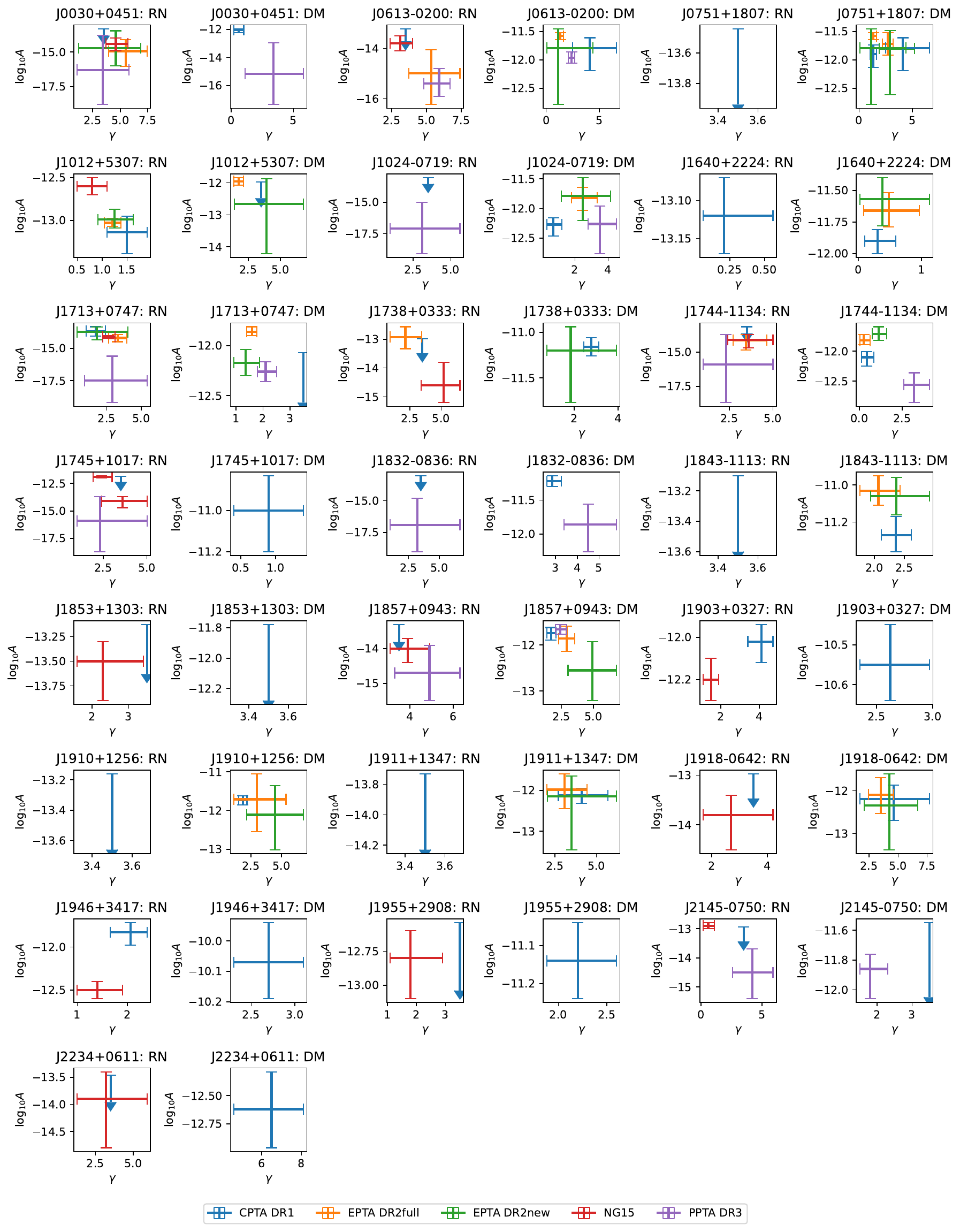}
\caption{Comparison of the red noise and DM GP power law parameters between the CPTA DR1, EPTA DR2full and DR2new, NANOGrav 15yr and PPTA DR3 data sets.}
\label{fig:comp}
\end{figure*}

\label{lastpage}

\end{document}